\documentclass{aa}  

\usepackage{graphicx}
\usepackage{txfonts}
\usepackage{natbib}
\usepackage{multirow}
\usepackage{amssymb}
\def\Rsolar{R$_{\odot}$}
\def\Msolar{M$_{\odot}$}
\def\perMsolar{$M_{\odot}^{-1}$}
\def\maa{$\alpha \alpha$}
\def\mga{$\gamma \alpha$}

\begin{document}

   \title{The demographics of neutron star - white dwarf mergers:  }

   \subtitle{rates, delay-time distributions, and progenitors }

   \author{S. Toonen\inst{1}\fnmsep\inst{2}\fnmsep \inst{3}
          \and
          H. B. Perets \inst{2}
          \and 
          A. P. Igoshev \inst{2} 
          \and
          E. Michaely \inst{2}
          \and
          Y. Zenati \inst{2}
          }

   \institute{Anton Pannekoek Institute for Astronomy, University of Amsterdam, 1090 GE Amsterdam, The Netherlands\\
              \email{toonen@uva.nl}
         \and Department of Physics, Technion, Haifa 3200003, Israel
         \and Departments of Physics and Astronomy, University of California, Berkeley, CA 94720, USA                }

   \date{Received April 5, 2018; accepted August 7, 2018}

\abstract
{The mergers of neutron stars (NSs) and white dwarfs (WDs) could give rise to explosive transients, potentially observable with current and future transient surveys. However, the expected properties and distribution of such events is not well understood. }
{Here we characterise the rates of such events, their delay-time distributions, their progenitors, and the distribution of their properties.}
{We use binary population synthesis models and consider a wide range of initial conditions and physical processes. In particular we consider different common-envelope evolution models and different NS natal kick distributions. We provide detailed predictions arising from each of the models considered. }
{We find that the majority of NS-WD mergers are born in systems in which mass-transfer played an important role, and the WD formed before the NS. For the majority of the mergers the WDs have a carbon-oxygen composition ($60-80\%$) and most of the rest are  with oxygen-neon WDs. The time-integrated rates of NS-WD mergers are in the range of $3-15\%$ of the type Ia supernovae (SNe) rate. Their delay-time distribution is very similar to that of type Ia SNe, but is slightly biased towards earlier times. They typically explode in young $100\,{\rm Myr}<\tau<1\,{\rm Gyr}$ environments, but have a tail distribution extending to long, gigayear-timescales. Models including significant kicks give rise to relatively wide offset distribution extending to hundreds of kiloparsecs. }
{The demographic and physical properties of NS-WD mergers suggest they are likely to be peculiar type Ic-like SNe, mostly exploding in late-type galaxies. Their overall properties could be related to a class of  recently observed rapidly evolving SNe, while they are less likely to be related to the class of Ca-rich SNe.}


   \maketitle

\section{Introduction}
\label{sec:intro}
  The mergers of double-compact-object binaries can give rise to explosive and/or transient events and the formation of exotic objects, which otherwise do not form through the evolution of single stars. Numerous studies have explored the mergers of double-white-dwarf (WD) binaries, mostly in the context of type Ia supernova (SN) progenitors \citep[e.g.][for reviews]{Liv18, Sok18, Wan18}. The mergers of black holes (BHs) and neutron stars (NSs) and their combinations (BH-BH, BH-NS, NS-NS) have also been explored extensively, both as potential gamma-ray burst (GRB) progenitors, and as gravitational wave (GW) sources such as those detected by the advanced LIGO-VIRGO consortium \citep{Abb16,Abb17}. Nevertheless, the mergers of other double-compact-object binaries, such as NS-WD binaries (which are likely the most common type of double-compact-object binaries besides WD-WD binaries; \citealt{Nel01c}) have received less attention.

\cite{Pas11} addressed the inspiral and merger of binary NS-WD with fully general relativistic simulations. They find that the merger remnant is a spinning Thorne-$\rm \dot{Z}$ytkow-like object surrounded by a massive disk and is undergoing a delayed collapse.
  \cite{Kin07} suggested that the accretion of the debris of the disrupted WD following the merger can give rise to a unique type of long gamma-ray burst. 
From a statistical point of view, \cite{Cha07} came to a similar conclusion.
Recently, \cite{Met12} studied the outcomes of NS-WD mergers and suggested the early phases of accretion give rise to faint thermonuclear explosions occurring in the accretion disk  \citep[see also][]{Mar16, Mar17}. Our recent study (Zenati et al., in prep.) further explored such mergers through more detailed models, finding that they are mostly driven by accretion with only very little contribution from thermonuclear sources, mostly consistent with the previous results. Such events may also contribute to the chemical evolution of galaxies, as their nucleosynthentic products somewhat differ from those of ordinary supernovae (SNe) \citep{Mar16, Mar17}. 
  
  Although NS-WD mergers may have various observable explosive outcomes, the demographics of such events and their progenitors have been little explored. Here we use extensive population synthesis models of binary stellar evolution to characterise the demographics and rates of NS-WD mergers. We note that NS binaries may also form in clusters and in the field through dynamical captures \citep{Pos14,Mic16b,Kle+17}; here we only consider the unperturbed evolution of isolated primordial binaries. 
  Given the many uncertainties involved in the modelling of binary evolution, we explore several different models, each differing from the others in its approach to the main uncertainties (e.g.  the distribution of NS natal-kicks, the properties of common envelope evolution, etc.). Our models provide us with detailed predictions for the overall rates of WD-NS mergers, their delay time distribution (DTD), expected host galaxies, and the distribution of the properties of their progenitors (e.g. WD masses and composition). In addition we briefly discuss the role of such mergers in producing GW sources observable by next-generation GW detectors.
  
    The last two decades of observational progress and the advent and development of large-scale automated SN surveys have revealed the existence of many novel types of "peculiar" transients, including several families of non-standard type Ia and other likely thermonuclear SNe \citep{Li01, Li11}. Future and upcoming surveys such as ZTF, ATLAS, and LSST are likely to discover many more peculiar transients, and in particular faint and/or fast-evolving transients not observable by current shallower/low-cadence surveys. If NS-WD mergers result in peculiar transients as predicted by current models, they might be observable over the coming decade. The characterisation of such events and their distribution is therefore a critical step in identifying them. Moreover, these studies open the window for connecting current theory with observations and constraining the models for NS-WD mergers, their outcomes, and even their potential role in affecting the chemical evolution of galaxies. 
  
  In the following we first discuss the population synthesis models that we employ and their various properties (Sect.\,\ref{sec:pop_syn}). We then describe our results regarding the evolution, rates, delay-time distribution, and galaxy offset of NS-WD mergers (Sect.\,\ref{sec:res}),   and finally discuss their implications and conclude (Sect.4). 
  
\section{Population synthesis}
\label{sec:pop_syn}
The formation and evolution of compact binary mergers is simulated with the binary population synthesis (BPS) code \texttt{SeBa} \citep{Por96, Too12, Too13}. \texttt{SeBa} is a fast code for simulating binary evolution based on parametrised stellar evolution, including processes such as stellar winds, mass transfer episodes, gravitational wave emission and supernova kicks. We employ \texttt{SeBa} to generate a large population of binaries on the zero-age main sequence (MS). We then simulate their subsequent evolution, and extract those that lead to a merger between a NS and a WD. 

It was shown in \cite{Too14} that the 
main sources of  differences between different BPS codes is due to the choice of input physics and initial conditions. To assess the systematic uncertainties in our predictions we construct different models that differ with respect to the SN kick and the common-envelope (CE) phase, as described  by the main models in Table\,\ref{tbl:rate}. Furthermore, we construct an additional set of models that differs with respect to the population of primordial binaries and assumptions regarding stable mass-transfer processes. 
For these additional models we assume a single distribution of SN kicks \citep[based on][Sect.\,\ref{sec:sn}]{Hob05} and consider two approaches for modelling the CE phase (model \maa\  and \maa2, since model \mga\,gives similar results as model \maa, see Sect.\,\ref{sec:ce}). 
The different models are described in the following sections.

In this paper, the terms primary  or secondary refer to the initially more or less massive component of a given binary, respectively.  

\subsection{Primordial binaries}
\label{sec:primordial}

In our standard models, the primordial binaries are generated as follows.\\
- Primary masses $M_1$ are drawn from a Kroupa IMF \citep{Kro93} with masses in the range 4-25\Msolar. For the normalisation of the rates, primary masses between 0.1 and 100\Msolar\,are considered. \\
- Secondary masses $M_2$ are drawn from a flat mass ratio distribution with $0<q\equiv M_2/M_1 < 1$ \citep{Rag10, Duc13}. \\
- The orbital separations $a$ follow a flat distribution in $log(a)$ \citep{Abt83}. \\
- The initial eccentricities $e$ follow a thermal distribution \citep{Heg75}. \\
- A constant binary fraction\footnote{
There are indications that the binary fraction increases with the mass of the primary star \citep{Rag10, Duc13,Moe17}. In Sect.\,\ref{sec:trans} we discuss the effect of a non-constant binary fraction on the NS-WD merger rate.} binary fraction $\mathcal{B}$ of 75\% is assumed which is appropriate for B-type primaries \citep{Rag10, Duc13,San14}. 

\subsection{Supernova kicks}
\label{sec:sn}
At the end of the life of a high-mass star, its core collapses under the pressure of self-gravity. 
During the core-collapse (CC) SN, a kick is imparted to the NS, as suggested by studies of pulsar scale heights \citep[e.g.][]{Gun70}, proper motions of pulsars \citep{Cor93,Hob05, Lyn94,Ver2017}, and the high velocities of some single NSs \citep{Cha05, Bec12}. 
Kicks solely due to the sudden mass loss are referred to as the 
`Blaauw-kicks' \citep{Bla61}, which can be limited due to interactions in close binary systems that strip the stellar envelopes of the donor stars prior to the SN \citep{Hua63, Tut73, Leo94}. 
The formation mechanism of additional `natal-kicks' is an unsolved problem \citep[see e.g.][]{Kus96, Sch06, Won13, Hol17, Kat18}, but likely involves anisotropies in the neutrino losses and/or in the mass loss in the SN ejecta \citep[][for a review]{Jan12}.

In this paper we adopt four different models for the SN kick. In our main model we randomly draw a natal kick from the distribution of \cite{Hob05} which is a Maxwellian distribution with a one-dimensional (1D) root mean square (rms) of $\sigma=265$km/s. As an alternative model we adopt the distribution of \cite{Arz02} which has two Maxwellian components, one at high velocities ($\sigma=500$km/s) and one at low velocities ($\sigma=90$km/s), as derived for high-mass X-ray binaries, double NS binaries, and pulsars retained in globular clusters \citep{Pfa02, Sch10, Ben16}. 
Recently from a direct comparison of pulsar parallaxes and proper motions, \citet{Ver2017} derived a kick distribution that exists of two Maxwellian components with $\sigma=316$km/s and $75$km/s.
The origin of the reduced natal kicks has been linked to different CC scenarios (i.e. electron-capture SN vs. iron CC, or accretion-induced collapse), or linked to the specific  masses of the core and ejecta \citep[][and references therein]{Pod04, Heu04,Kni11, Jan12, Tau15,Bra16, Jan17}. 
As a lower limit on the SN kicks, we constructed a third model without natal kicks, and so the SN-kick consists of only the effect of the instantaneous mass loss, that is, a Blaauw-kick \citep{Bla61}. 

\subsection{The common-envelope phase}
\label{sec:ce}
Common-envelope evolution is a mass-loss phase in the formation of compact binaries, and is thought to lead to a severe shrinkage of the binary orbit, thus explaining the existence of compact binaries ($a\lesssim 1-10$\Rsolar) containing one or two compact objects whose progenitors would not fit in that orbit during the giant phases. However, despite the importance of the CE phase for many types of binaries (e.g. supernova type Ia progenitors, cataclysmic variables, low-mass X-ray binaries), and the tremendous effort of the community, the CE phase is poorly understood \citep[for a comprehensive review, see][]{Iva13}. 

The CE phase is commonly modelled based on the conservation of energy \citep{Pac76, Web84, Liv88, DeK87, DeK90}:
\begin{equation}
E_{\rm bin} = \frac{GM_dM_c}{\lambda R} = \alpha_{\rm CE} E_{\rm orbit},
\end{equation}
where $M_d$ is the mass of the donor star, $M_c$ the mass of its core, $R$ its radius, $\lambda$ the structure parameter of its envelope, and $E_{\rm bin}$ is the binding energy of the envelope. As a source of energy to unbind the envelope, classically the orbital energy $E_{\rm orbit}$ is considered, with an efficiency of $\alpha_{\rm CE}$ \citep[but see e.g.][]{Sok04, Iva15, Gla18}. 

In this paper we focus on the classical energy balance of CE evolution, and vary the efficiency with which orbital energy can be used to expel the donor's envelope. In model \maa\, we assume $\alpha_{\rm CE}\lambda=2$, whereas for model \maa2,  $\alpha_{\rm CE}\lambda=0.25$ is assumed.
The prior is calibrated to formation of double WDs \citep[i.e. second phase of mass transfer, see][]{Nel00, Nel01}, whereas the latter is calibrated on the formation of compact WD-MS systems where the MS is of spectral type M \citep{Zor10, Too13, Cam14, Zor14}. 

An alternative model for CE evolution is the $\gamma$-CE which is based on a balance of angular momentum instead of energy, as 
\begin{equation}
\frac{J_{\rm init}-J_{\rm final}}{J_{\rm init}} = \gamma \frac{\Delta M_d}{M_d+ M_a},
\end{equation} 
where $J_{\rm init}$ and $J_{\rm final}$ are the angular momentum of the pre- and post-mass-transfer binary, respectively, and $M_a$ is the mass of the companion. The $\gamma$-prescription was introduced to explain the first phase of mass transfer in the formation of double WDs \citep{Nel00,Slu06}. In our model \mga\,, we apply the $\gamma$-CE with $\gamma=1.75$, unless the companion is a degenerate object or the mass transfer is dynamically unstable, for which the $\alpha$-CE with $\alpha\lambda=2$ is applied \citep{Nel01}. 

\subsection{Stable mass transfer}
\label{sec:mt3}

Stable mass transfer between two hydrogen-rich stars is a frequent phenomenon in the evolution of the NS-WD merger progenitors (see also Sect.\,\ref{sec:ev}). Therefore, it is important to assess how our standard assumptions impact the synthetic rates. We construct three models, which differ in several aspects.
\begin{itemize}
\item \textbf{The angular-momentum loss mode} \citep{Pol94, Sob97,Too14}. When the mass transfer is not completely conservative, not only mass but also angular momentum is lost from the system. The effect of this on the orbit can be severe, and depends crucially on how the mass is lost; for example, whether the mass is lost directly from the primary, or whether is first crosses into the Roche lobe of the secondary, and subsequently is expelled from close to the surface of the secondary. The standard assumption in \texttt{SeBa} when the accretor is a non-degenerate star is that the specific angular momentum of the lost matter is 2.5 times the specific angular momentum of the orbit \citep{Por95, Nel01}. 
When the accretor is a compact object and mass transfer is non-conservative, it is assumed that the lost matter has the specific angular momentum of that of the compact object. This mode of angular momentum loss is also known as `isotropic re-emission'. Here we test the effect of the standard assumption in \texttt{SeBa}; in an alternative model we assume that the specific angular momentum is equal to that of the secondary for all types of accretors. This model is motivated by the work of \cite{Woo12} who demonstrated that stable mass transfer can be more readily realised under this assumption for stars of about one solar mass in the context of the formation of double white-dwarfs.

\item \textbf{The stability of mass transfer} is dependent on the reaction of the stellar radii $R$ and the corresponding Roche lobes $R_{\rm RL}$ to the transfer of mass and angular momentum. When mass transfer proceeds on timescales roughly equal to or shorter than the thermal timescale of the donor star, the donor can no longer achieve thermal equilibrium (or possibly even hydrostatic equilibrium). Modelling of such mass-transfer phases and determining the stability of mass transfer for different types of binaries is not trivial. Regarding BPS codes, as they rely on stellar evolution models of stars in thermal and hydrostatic equilibrium, they cannot adequately calculate the donor properties and the stability of mass transfer. Instead, BPS codes rely on parametrisations or
interpolations to determine the stability of mass transfer. These are based on simplified stellar models (e.g. polytropes) or calculations from detailed stellar evolution codes \citep[e.g.][]{Hje87,deM07,Ge10,Ge15}. It is important to realise that even the detailed stellar evolution codes are not adequate in simulating mass transfer on timescales sufficiently above the donor's thermal timescale \citep[e.g.][]{Pav15}. Nonetheless, these types of studies have shown that stable mass transfer in binaries with giant donors is more readily realised than previously assumed \citep[e.g.][]{Pav15, Pav17}, for example, due to the short local thermal timescale of the super-adiabatic outer layer of a giant's envelope \citep{Woo11, Pas12}\footnote{Enhancements of the stability of mass transfer in systems with giant donors with shallow convective envelopes is incorporated in \texttt{SeBa}, see Appendix A.3 of \citep{Too12}, and \cite{Too14}.}. 

As an alternative model, we test the sensitivity of   NS-WD mergers to our assumptions regarding the stability of mass transfer. We do this by adjusting the stability criterion in \texttt{SeBa}. This is based on the adiabatic response of the star:
\begin{equation}
\zeta_{\rm ad} \equiv \frac{d\ \rm  ln R}{d\ \rm ln M}, 
\label{eq:dlnrdlnm_star}
\end{equation} 
and 
\begin{equation}
\zeta_{\rm RL} \equiv \frac{d\ \rm  ln R_{\rm RL}}{d\ \rm ln M}, 
\label{eq:dlnrdlnm_rl}
\end{equation} 
of the Roche lobe. 
If $\zeta_{\rm RL} < \zeta_{\rm ad}$ we assume mass transfer proceeds in a stable manner \citep[e.g.][]{Web85, Pol94}. 
For every Roche-lobe-filling system, $\zeta_{\rm RL}$ is calculated numerically by transferring a test mass of $10^{-5}$\Msolar \ at every time-step. The value of $\zeta_{\rm ad}$ depends on the type of star considered and is  tabulated in Appendix A3 of \citet{Too12}. In the alternative model, we decrease $\zeta_{\rm ad}$ from 4 to 2 for all stellar types with radiative or shallow convective envelopes. 

\item \textbf{The accretion efficiency} $\beta$ is 
the degree to which mass transfer is conservative, that is
the fraction of mass lost by the donor star that can be accreted by the companion (i.e. $\beta\equiv \dot{M}_d/\dot{M}_a$). 
Modelling of the evolution of some binaries (such as $\phi$ Per) indicates that mass transfer can proceed fairly conservatively \citep{Pol07}, however for other systems, spin-up of the accretor is expected to limit the amount of accretion \citep[e.g.][]{Pac81}. Attempts to constrain the efficiency have so far remained inconclusive \citep[e.g.][and references therein]{deM07}. 

As the viability of some of the evolutionary channels (i.e. `semi-reversed WDNS' and `reversed NSWD', see Sect.\,\ref{sec:ev}) is directly related to the amount of accretion onto the secondary, we constructed two alternative models
that assume completely conservative and 50\% conservative mass transfer onto non-degenerate companion stars (i.e. $\beta=1$ and $\beta=0.5$), respectively. 
These models are mathematical exercises to assess the robustness of the predicted rates. 
In comparison, in the standard models we assume that the accretion rate onto these stars is limited to a factor times the thermal timescale of the accretor. The factor is dependent on the ratio of Roche-lobe radius to the effective radius of the accretor star \citep[][see also Eq. C13 in \cite{Por96}]{Pol94, Por96, Too12}. This inhibits accretion in systems with low mass ratios $q = M_2/M_1$.
\end{itemize}

\section{Results}
\label{sec:res}

\subsection{Evolution}
\label{sec:ev}

In this section, we describe the evolutionary path for progenitors of binary mergers between a NS and a WD. In Sect.\,\ref{sec:wdns_direct}-\ref{sec:nswd_rev} we discuss the formation of the NS-WD binary, whereas Sect.\,\ref{sec:towards_merger} focuses on the subsequent evolution leading to the merger. We identify the four main channels, of which the first is
a dominant channel for most models and consistently a major contributor (Table \,\ref{tbl:rate}).

We distinguish between systems in which the NS forms first (hereafter NSWD), and where the WD forms first (hereafter WDNS). When we refer to all mergers, combining those of NSWD and WDNS, we use the term NS-WD mergers.

\subsubsection{Pathway 1: direct WDNS}
\label{sec:wdns_direct}

\begin{table}
\caption{Definitions of abbreviations of stellar types used in the text and figures.}
\begin{tabular}{|l|l|}\hline 
Abbreviation &  Type of star  \\
 \hline  \hline 
MS & Main sequence star\\
HG & Hertzsprung-gap star\\
He MS & Star on the equivalent of the main sequence \\
 & for hydrogen-poor helium-burning stars \\
He G & Hydrogen-poor helium-burning giant \\
WD & White dwarf \\
NS & Neutron star \\
BH & Black hole \\
SN & Supernova explosion \\
 \hline 
 \end{tabular}
\label{tbl:star_type}
\end{table}

\begin{figure}
\centering
\includegraphics[width=\columnwidth, clip=true, trim =0mm 50mm 0mm 30mm]{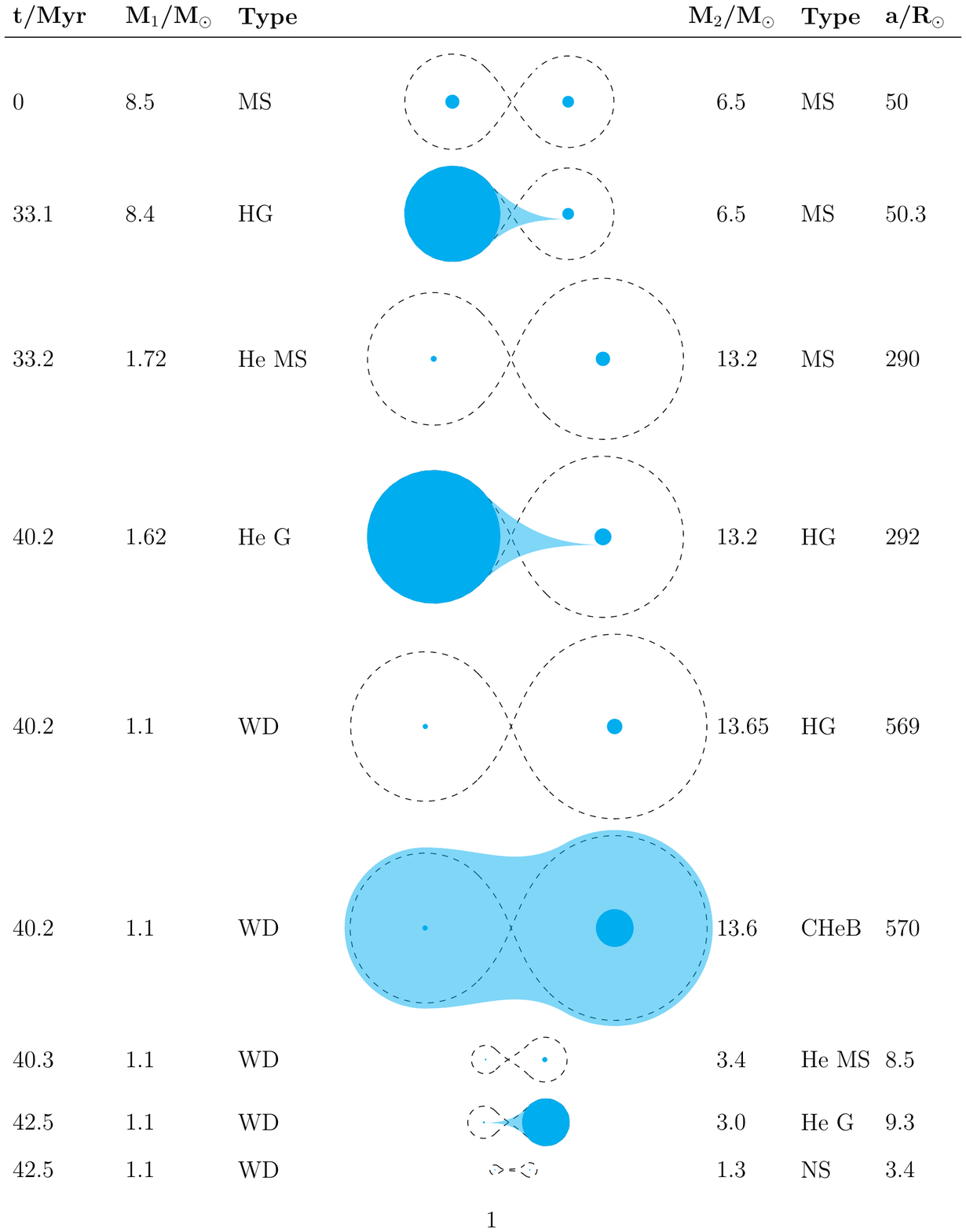} 
\caption{Example of the evolution of a system in the `direct WDNS' channel. 
Orbital and stellar parameters are given in the columns beside the illustration of the binary. The abbreviations of the stellar types are defined in Table\ref{tbl:star_type}. The extent of the Roche lobes and stars in the illustration are approximate and are not drawn to scale. } 
\label{fig:wdns_direct}
\end{figure}

The primordial binaries of this channel have primary and secondary masses in the range of $\sim$7-11\Msolar, and $\sim$4.5-11\Msolar, respectively. The initial orbits are relatively small; semi-latus recti in the range $a_{\rm SLR} = a(1-e^2)\sim 20-100$\Rsolar. The primary star fills its Roche lobe during the Hertzsprung gap, giving rise to a phase of stable mass transfer, and once more as a hydrogen-poor helium-rich star of about 1-2.3\Msolar, before turning into a WD.
During the prior mass-transfer phases, the secondary has accreted a significant amount of mass, rendering it possible to collapse to a NS at the end of its life and making the orbit of the WDNS eccentric. 

We find that the most important channel of NS-WD mergers is the one in which the WD is formed before the NS (WDNS channel). Due to the highly conservative mass transfer onto the secondary, 
the secondary becomes massive enough to undergo a SN explosion and form a NS. 
The possibility of a reversal of the end states of the two components has previously been noted, for example for NS-NS \citep{Por96}, NS-BH \citep{Sip04} and NS-WD binaries \citep{Tut93,Por99, Tau00, Dav02, Chu06}.
This is observationally supported by a few binary pulsars with WD companions \citep{Por99, Tau00, Dav02, Chu06}; PSR B2303+46  \citep{Ker99}, PSR J1141-6545 \citep{Kas00,Man00}, and PSR B1820-11 \citep{Lyn89}, and likely also PSR J1755-2550 \citep{Ng18}. These systems contain young NSs (i.e. non-recycled pulsars with large magnetic fields 
and short spin-down times) in eccentric orbits \citep[see also][]{Zha11}.

\subsubsection{Pathway 2: direct NSWD}
\label{sec:nswd_direct}

\begin{figure}
\centering
\includegraphics[width=\columnwidth, clip=true, trim =0mm 50mm 0mm 30mm]{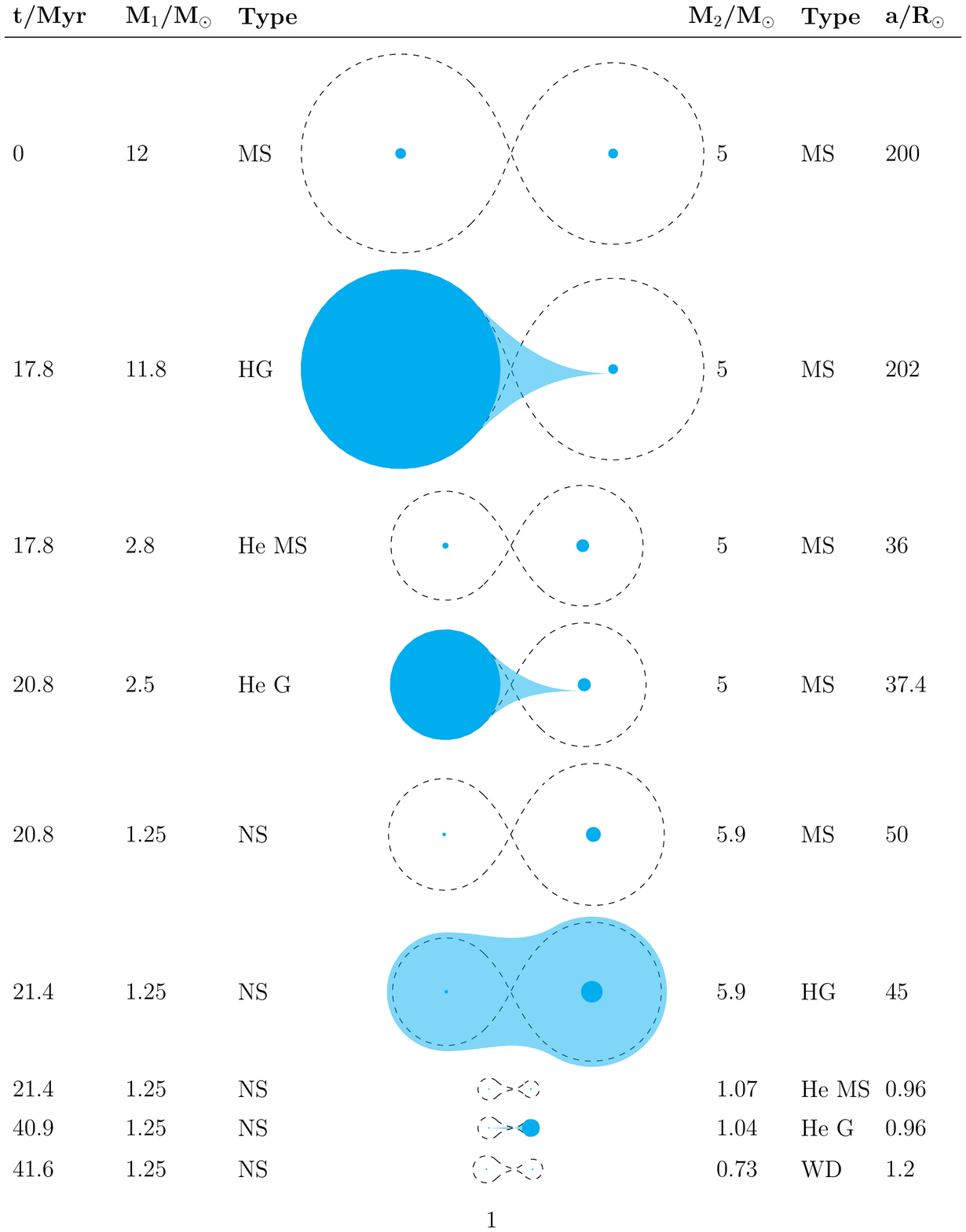} 
\caption{
Example of the evolution of a system in the 'direct NSWD'-channel. 
The layout and legend are similar to those in Fig.\,\ref{fig:wdns_direct}. } 
\label{fig:nswd_direct}
\end{figure}

The initial stellar masses and orbital separations are larger compared to the previous channel. The primary masses are mostly in the range $\sim$11-19\Msolar\, (but extending down to 8\Msolar), the secondary masses are in the range $\sim$2-10\Msolar, and the semi-latus rectii are in the range $a_{SLR}\sim 50-2000$\Rsolar.

For these systems the primary evolves into a NS first. Subsequently the secondary star initiates the mass-transfer phases, loses its hydrogen and helium envelope, and ends its life as a WD. An example for such evolution is shown in Fig.\,\ref{fig:nswd_direct}. During the mass-transfer phases after the formation of the NS, the system 
could be observed as an X-ray binary \citep[e.g.][]{Tau06}. Any accretion of mass and angular momentum \citep{Lor08,Tau15b,Man17} would likely spin up  or `recycle' the NS, suppress the magnetic field of the NS, and circularise the orbit  \citep{Bis74, Alp82}.

\subsubsection{Pathway 3: semi-reversed WDNS}
\label{sec:wdns_semirev}

\begin{figure}
\centering
\includegraphics[width=\columnwidth, clip=true, trim =0mm 70mm 0mm 45mm]{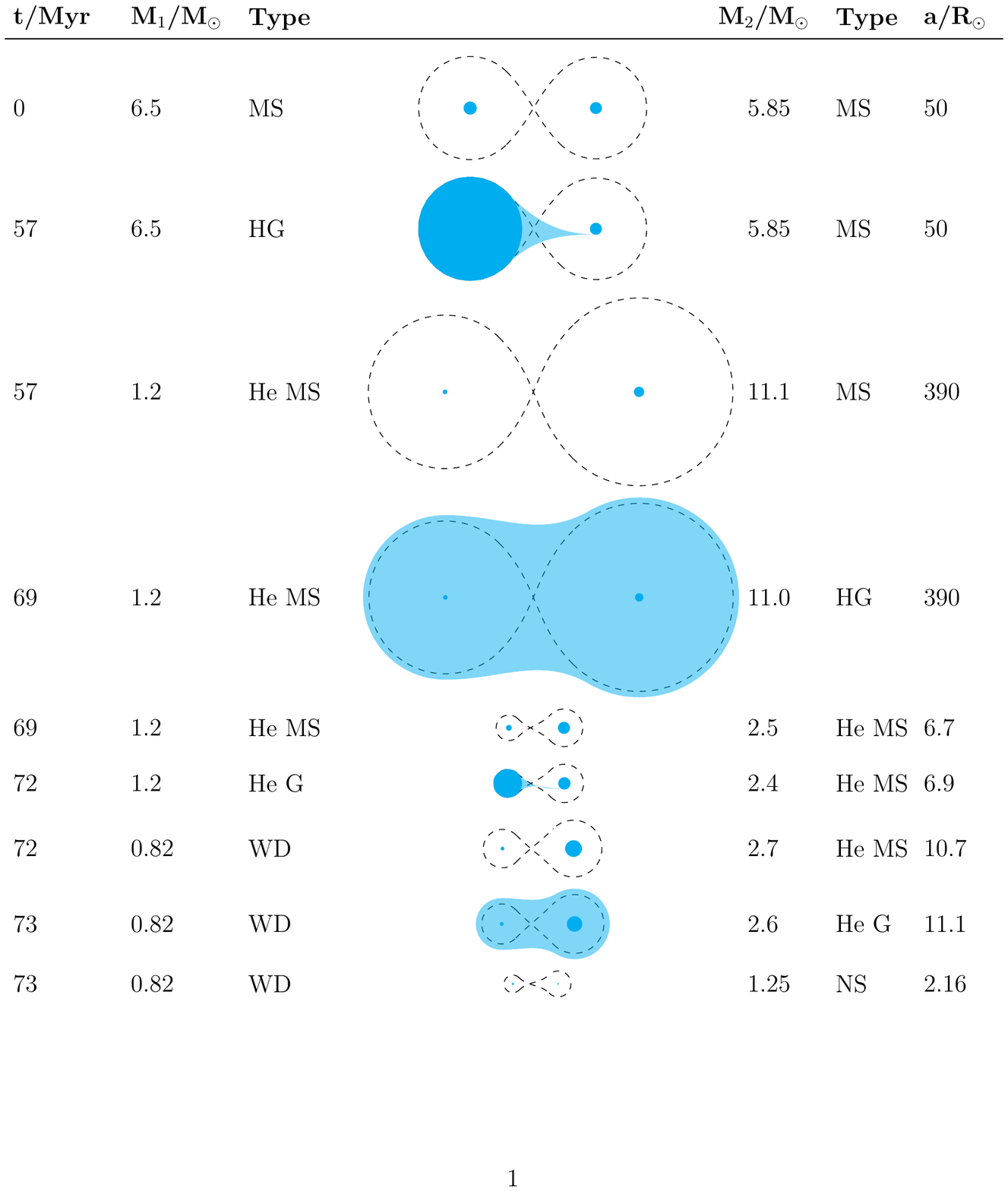} 
\caption{Example of the evolution of a system in the 'semi-reversed WDNS'-channel. 
In this channel there is a stage where both stars have been stripped of their hydrogen envelopes before forming a compact object.
The layout and legend is similar to that of Fig.\,\ref{fig:wdns_direct}. } 
\label{fig:wdns_semirev}
\end{figure}

In this channel (Fig.\,\ref{fig:wdns_semirev}.) the primary becomes a WD before the secondary becomes a NS, hence the name WDNS. The terminology `semi-reversed' reflects the neck-to-neck race that the stars are in to form the first compact object in the system. The secondary has already initiated a phase of mass transfer and evolved to the hydrogen-poor helium-burning tracks before the WD primary is formed. 

In this evolutionary channel, the initial primary masses and semi-latus rectii are similar to those of channel 1 (direct WDNS), however the initial mass ratios are preferably close to one, $q_i\sim0.85-1.0$. 
As the initial stellar masses and evolutionary timescales are similar, the secondary accretes a significant amount of mass during the first mass-transfer phase. 
As a result, its evolution is accelerated, the secondary expands and fills its Roche lobe while the primary is still a helium star. 
Similar to the `direct WDNS' channel, the orbit of the WDNS is eccentric at its formation.

\subsubsection{Pathway 4: reversed NSWD}
\label{sec:nswd_rev}

\begin{figure}
\centering
\includegraphics[width=\columnwidth, clip=true, trim =0mm 80mm 0mm 60mm]{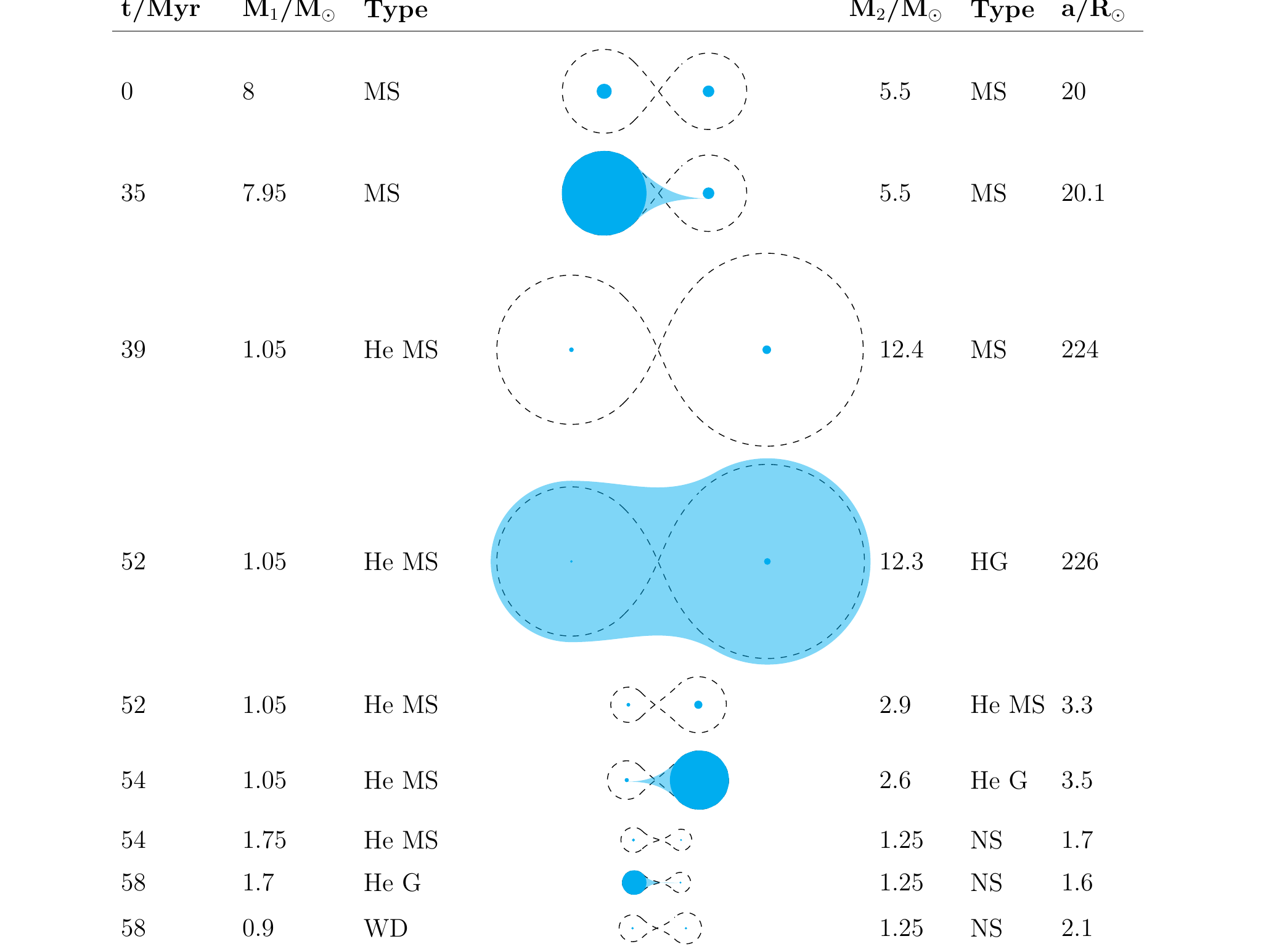} 
\caption{Example of the evolution of a system in the `reversed NSWD' channel. 
The secondary collapses to a NS before the primary becomes a WD. The evolution of the stars is therefore `reversed'. The layout and legend are similar to that in Fig.\,\ref{fig:wdns_direct}. } 
\label{fig:nswd_rev}
\end{figure}
 The evolution of this channel is shown in Fig.\,\ref{fig:nswd_rev}.
 The NS is formed before the WD, similar to channel 2 (direct NSWD). In the `reversed NSWD' channel, the evolution of the secondary typically overtakes that of the primary, and the secondary becomes the first compact object in the binary. 
 The detailed evolution is similar to that of channel 3 (semi-reversed WDNS), with the order of the third and fourth mass-transfer phases being reversed. Due to the strong accretion, the subsequent evolution of the secondary, both as a hydrogen-rich and helium-rich star, is shorter than that of the helium-rich primary. 

We note that this evolutionary pathway shows similarities to that described in \citet[][`Formation reversal channel']{Too12} for the formation of double WD mergers as progenitors of Type Ia supernovae.

The initial masses of both stars are approximately in the range of 5-8\Msolar\,in this channel. The initial mass ratios are predominantly between 0.85 and 1, but they can extend up to 0.5. The initial semi-latus rectii $a_{\rm SLR}$ are   $\lesssim 200$\Rsolar). 

\subsubsection{Evolution towards the merger}
\label{sec:towards_merger}

After the formation of the WDNS system, the orbit shrinks due to the emission of gravitational waves (GWs). 
If the orbit is eccentric (such as for the WD-NS binaries), the GW in-spiral time $t_{\rm insp}$ can be significantly reduced as given by \citep{Pet64}:

\begin{equation}
t_{\rm insp} = \frac{ 60c^5c_0^4}{1216G^3m_1m_2(m_1+m_2)} \int_{0}^{e_0} de \frac{e^{29/19} \Big[1+\frac{121}{304}e^2\Big]^{1181/2299}}{(1-e^2)^{3/2}},
\end{equation}
where $c_0 = \frac{a_0(1-e_0^2)}{e_0^{12/19}}\Big[1+\frac{121}{304}e_0^2\Big]^{-870/2299},$ 

with $a_0$ and $e_0$ being the initial semi-major axis and eccentricity.
For example, a 1.3\Msolar\, NS and a 0.8\Msolar\, WD in a circular orbit of 1\Rsolar\, merge after about 70 Myr, whereas an imposed eccentricity of 0.5 (0.75) at the birth of the NS reduces the in-spiral time to 25 (14) Myr.

Once the orbital period has shrunk to a few minutes, the WD fills its Roche lobe. If the subsequent mass transfer proceeds stably, an ultra-compact X-ray binary (UCXB) forms. The time-averaged X-ray luminosity of the majority of UCXBs is consistent with that expected from WD donors \citep{Hei13}, although other types of donor stars have also been suggested, such as helium burning stars or evolved MS stars \citep[see e.g.][for a population synthesis study]{Haa13}.
If unstable mass transfer develops in the NS-WD system, the WD is quickly disrupted on a dynamical timescale, leading to a merger of the system.

Studies of the stability of mass transfer from a WD to a NS typically discuss the critical mass ratio or critical WD mass $M_{\rm WD, crit}$ above which Roche-lobe overflow leads to a merger. Depending on the structure, composition and temperature of the WD and the mode and amount of mass and angular momentum loss,  $M_{\rm WD, crit} \sim 0.4-0.55$\Msolar\,\citep[e.g.][]{Ver88, Pas09, Yun02, Haa13}. 
 However, during the early stages of the mass transfer, its rate is often highly super-Eddington, such that disc winds become important. \cite{Bob17} recently showed that the disc winds reduce the stability of the mass transfer. They determine a critical WD mass of $M_{\rm WD, crit} =0.2$\Msolar, excluding all carbon-oxygen and oxygen-neon WDs as UCXB donor stars.

Assuming the stability limit of \citep{Haa13} ($M_{\rm WD, crit} = 0.38$\Msolar based on isotropic re-emission for the loss of angular momentum), over 97\% of semi-detached NS-WD systems merge in all models, whereas for $M_{\rm WD, crit} = 0.2$\Msolar~ over 99.9\% merge. In this work we adopt the latter prescription from \cite{Bob17}.

\subsection{NS-WD merger rates}
\label{sec:trans}

We find that the time-integrated rate of NS-WD mergers is  $(3-7)\cdot 10^{-5}$ per solar mass of created stars in most models (Table\,\ref{tbl:rate}).
We do not find significant differences in the merger rates between the different models of the CE phase, the stable mass-transfer phase, or the SN kick distributions of \cite{Hob05}, \cite{Arz02}, and \cite{Ver2017}. 
If all NS progenitors in interacting binaries receive minimal SN kicks (i.e. zero-velocity natal kicks), the time-integrated NS-WD merger rate increases by atmost a factor of 2-4 up to $(10-20) \cdot 10^{-5}$\perMsolar.
On the other hand the minimum total merger rate ($(0.3-3) \cdot 10^{-5}$\perMsolar) is achieved if the accretion efficiency of mass transfer is artificially reduced to a constant 50\%.
We note that model \mga\, gives very similar results to model \maa\, as the first phase of mass transfer tends to be stable for the systems considered in this paper (Sect.\,\ref{sec:ev}), and therefore model \mga\, is not considered for the additional models in Table\,\ref{tbl:rate}.

In this work, we assume a constant binary fraction $\mathcal{B}$ of 75\%, however there are indications that the binary fraction increases with the mass of the primary star \citep{Rag10, Duc13,Moe17}.  As a test, we follow \citet{Haa13} in assuming:
\begin{equation}
\mathcal{B}(M_1) = \frac{1}{2} + \frac{1}{4}\rm log_{10}(M_1).
\end{equation}
In this case, the time-integrated NS-WD merger rate only slightly increases; with  6-7.5\% higher rates across all models. 

We have also tested the effect of a different initial period and mass-ratio distribution. Firstly, instead of a uniform distribution of periods, we adopt a log-normal distribution of periods (in days) with parameters $\mu=4$ and $\sigma=1.3$ \citep{Duc13}. The main difference is that the number of short-period binaries (such as the progenitors in the first pathway `direct WDNS') is suppressed  compared to our main BPS models. Secondly, we adopt a mass-ratio distribution $N(q)$ that is strongly biased towards low-mass companions, that is $N(q) \propto q^{-2}$. This is motivated by observations of wide binaries with massive primaries \citep{Moe17}.
With these changes in the primordial binary population, the time-integrated NS-WD merger rate decreases by a factor of 4-5.  

Table\,\ref{tbl:rate} also shows the time-integrated NS-WD merger rate in each of the evolutionary channels identified in Sect.\,\ref{sec:ev}. Channel `direct WDNS' contributes to the total merger rate in a significant and consistent way across all models. The corresponding rate is $(2-5) \cdot 10^{-5}$\perMsolar\,in most models. The contribution from channel `direct NSWD' is strongly dependent on the CE modeling and the SN kick. The rate varies over two orders of magnitude, and so the strong increase in the total merger rate for smaller SN kicks can be attributed to this channel. 

The reason that the `direct NSWD' channel is more sensitive to the SN kick compared with the `direct WDNS' channel is related to the orbital separation of the pre-SN binary. In the latter channel, the pre-SN semi-major axes are several solar radii, but in the former they are several tens to hundreds of solar radii. This is because the systems in the `direct NSWD' channel have widened in response to the (stable) transfer of mass, whereas the other systems have undergone several mass-transfer phases and further common-envelope evolution in which the orbit has contracted 
(see Figs.\,\ref{fig:wdns_direct}\,and\,\ref{fig:nswd_direct}) \footnote{While in the `direct NSWD' channel, the rates increase when the average SN kick is lowered (e.g. going from the kick distribution of Hobbs to Arzoumanian), one notices that the rates of the `direct WDNS' channel are slightly higher given the Hobbs distribution compared to the Arzoumanian one. We attribute this to the high-velocity tail of the Arzoumanian distribution. Overall, the total merger rate is lowest for SN kicks following the Hobbs distribution, with the exception of model \maa2, where the contribution of the `direct NSWD' channel is low.}.

The third channel (`semi-reversed WDNS') provides $\sim$5-20\% of all mergers, and the `reversed NSWD'-channel gives rise to up to $\sim$10\% of the mergers. An exception to this is found in the additional \maa\,models regarding the stability of mass transfer and the angular momentum loss mode (see Sect.\,\ref{sec:mt3}). The enhanced accretion onto the secondary makes it possible for more secondaries to overtake the primary during its evolutionary progress. 
As a result, the `reversed NSWD' channel gives rise to $\sim$20\% of all mergers in the models with enhanced mass-transfer stability and isotropic re-emission of angular momentum. On the other hand, in our additional model with reduced accretion, the time-integrated rate of the `reversed NSWD' channel, as well as that of the `semi-reversed WDNS' channel are negligible.

\begin{table*}[h]
\caption{The time-integrated merger rate of NS-WD binaries per $10^5$\Msolar\,of created stars for the different BPS models (Sect. \ref{sec:pop_syn}). The main models vary with respect to the SN kick (drawn from the distribution of \cite{Hob05}, the distribution of \cite{Arz02}, or only the mass-loss kick, see e.g. \cite{Bla61}) and the common-envelope phase (model \maa, \maa2, or \mga). In the additional models several assumptions are varied in regards to the stable mass-transfer process.  
Columns 5-7 give the fraction of mergers with CO, ONe or He WDs, respectively. Columns 8-11 give the time-integrated merger rates from the four most common evolutionary channels described in Sect.\,\ref{sec:ev}.}
\begin{tabular}{|cll|c|ccc|cccc|}
\hline
\multicolumn{3}{|c|}{BPS Model} & \multicolumn{8}{|c|}{Rates ($10^{-5}$\perMsolar)}\\
\hline
\multirow{2}{*}{Type} & \multirow{2}{*}{CE} & \multirow{2}{*}{SN-kick} & \multirow{2}{*}{Total}  & CO WD  & ONe WD &  He WD  & direct & direct & semi-rev & rev \\
& &  &  & (\%)&  (\%) &  (\%) & WDNS& NSWD& WDNS& NSWD\\
\hline
\hline
\multicolumn{3}{|c|}{Main models  } & & & & & & & & \\
\hline
& \maa & Hobbs  &4.7 & 80&20 &0.4  & 2.7& 0.76& 0.77&0.53\\
 & \maa & Arzoumanian  & 5.6 & 84& 15 & 1.4 &2.2 & 2.1& 0.75&0.44\\
& \maa & Verbunt  & 7.4 & 84& 15 & 1.4 &2.9 & 2.9& 0.93&0.57\\
& \maa & Blaauw & 18& 89& 9& 2 & 3.3& 12.7&1.2 &0.72\\
\hline
& \maa2 & Hobbs  & 4.0& 63& 37 & -  & 3.2& 0.01& 0.65&0.15\\
 & \maa2 & Arzoumanian  &3.8 & 61& 39 & 0.07 & 2.9&0.10 &0.58 &0.15\\
& \maa2 & Verbunt  & 4.7 & 62& 38 & 0.09 &3.8 & 0.14& 0.67&0.16\\
& \maa2 & Blaauw & 8.1 & 66& 33& 0.4 & 5.1& 2.0& 0.74&0.25\\
\hline
& \mga & Hobbs &  3.7 & 77 & 22 & 0.7 & 2.8& 0.56& 0.24& 0.09\\
 & \mga & Arzoumanian & 4.0 & 83& 16 & 1.3 & 2.3& 1.5& 0.20& 0.03\\
& \mga & Verbunt  & 5.4 & 82& 16 & 1.4 &3.0 & 2.1& 0.29&0.06\\
& \mga & Blaauw & 12 &90 & 7& 2& 3.3& 8.1&0.30 & -\\
\hline
\hline
\multicolumn{3}{|c|}{Additional models  } & & & & & & & & \\
\hline
\hline
Isotropic & \maa & Hobbs & 5.7 & 81&19& 0.5 & 3.0& 0.49& 1.0&1.2\\
re-emission & \maa2 & Hobbs  &4.4 & 64& 36& - & 3.4& 0.02& 0.90&0.14\\
\hline
Mass transfer &  \maa & Hobbs & 5.6 & 81&19 & 0.6 & 2.8&0.66 &0.92 &1.2\\
stability &  \maa2 & Hobbs & 4.2 & 63& 37& - & 3.2& 0.01&0.92 &0.13\\
\hline
Conservative & \maa & Hobbs & 4.8 & 80&20 &0.6  & 2.8& 0.77& 0.80&0.45\\
mass transfer & \maa2 & Hobbs & 4.8 & 62& 38& - & 4.0& 0.002& 0.67&0.16\\
\hline
50\% conservative & \maa & Hobbs & 3.2 & 43&57 &-  & 2.5& 0.73& - &-\\
mass transfer & \maa2 & Hobbs & 0.26 & 42& 58& - & 0.24& 0.02& -& - \\
\hline
\hline
\end{tabular}
\label{tbl:rate}
\end{table*}

\subsubsection{Delay-time distributions}
\label{sec:dtd}
\begin{figure}
    \centering
        \includegraphics[width=\columnwidth]{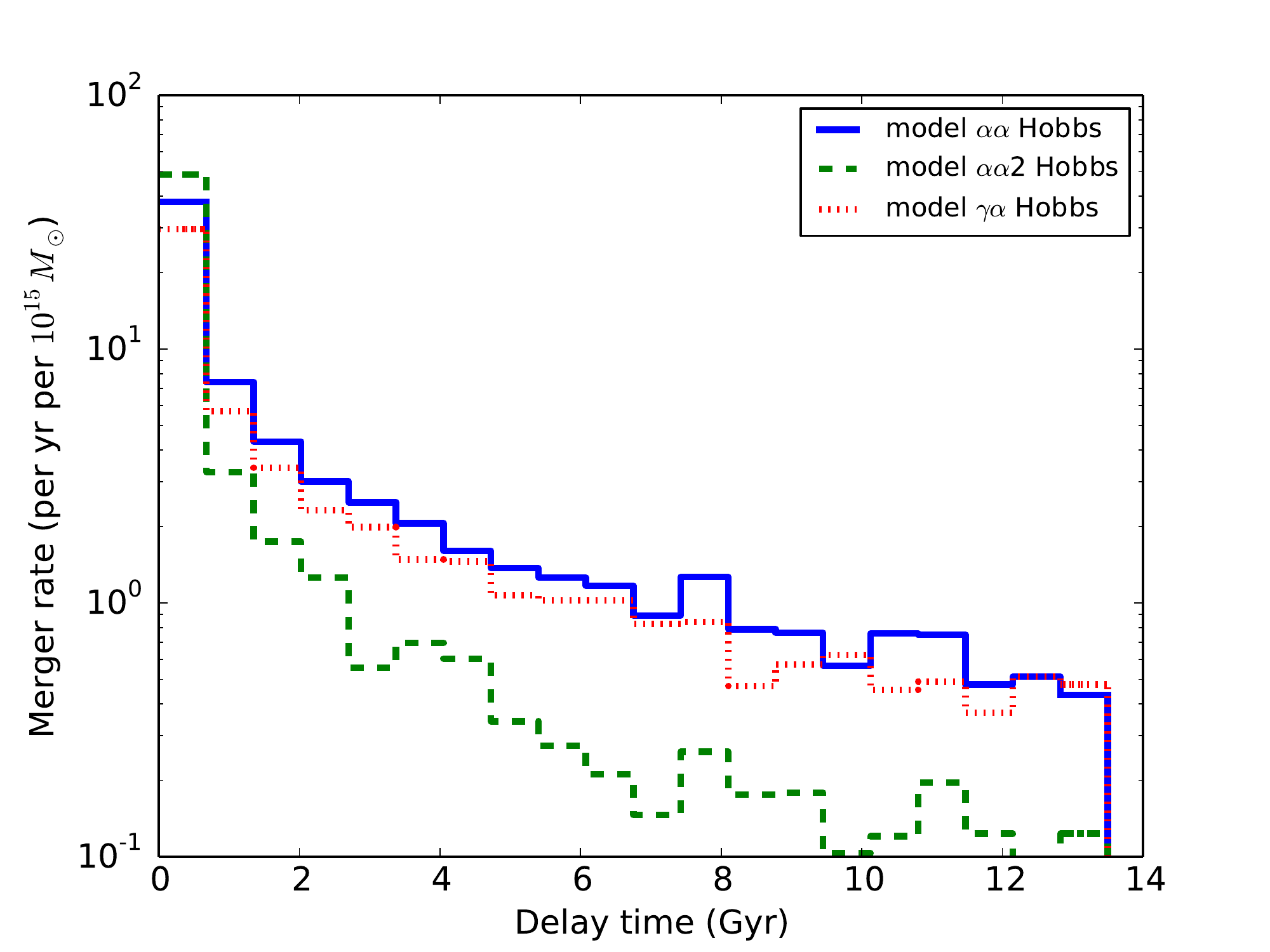}
    \caption{Effect of CE evolution on the delay-time distributions of NS-WD mergers. The figure shows DTDs for different models of the common-envelope phase with SN-kicks that follow the Hobbs distribution. The slope of the DTD is strongly dependent on the CE efficiency. Model \maa\, ($\alpha$-CE with $\alpha\lambda=2$, solid blue line) and model \mga\, ($\gamma$-CE, dotted red line) give similar DTDs with tails extending to long delay times. Mergers at long delay times are suppressed in model \maa2\, ($\alpha$-CE with $\alpha\lambda=0.25$, dashed green line).} 
    \label{fig:DTD_CE}
    \end{figure}

\begin{figure}
    \centering
    \begin{tabular}{c}
        \includegraphics[width=\columnwidth]{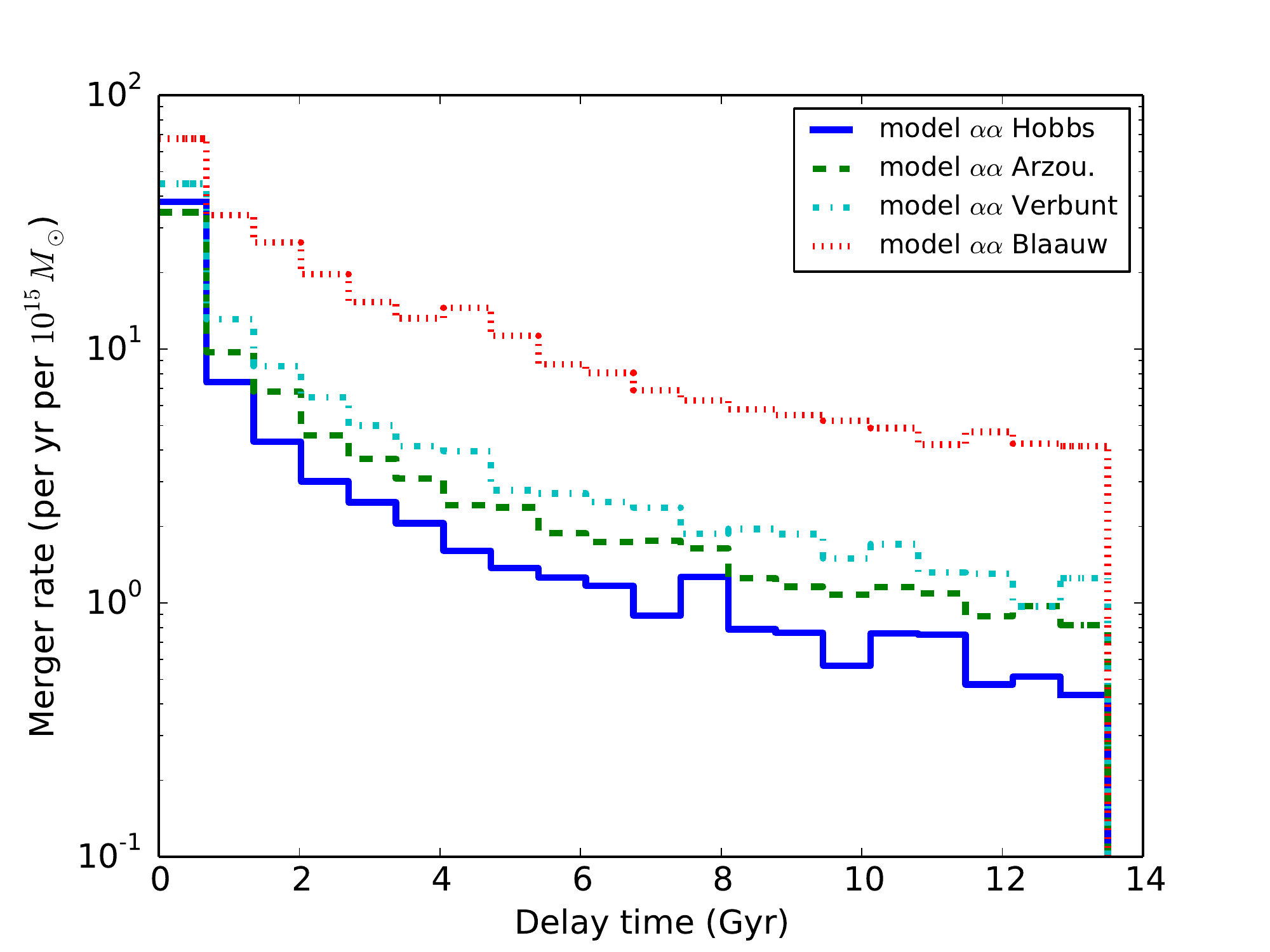}  \\
        \includegraphics[width=\columnwidth]{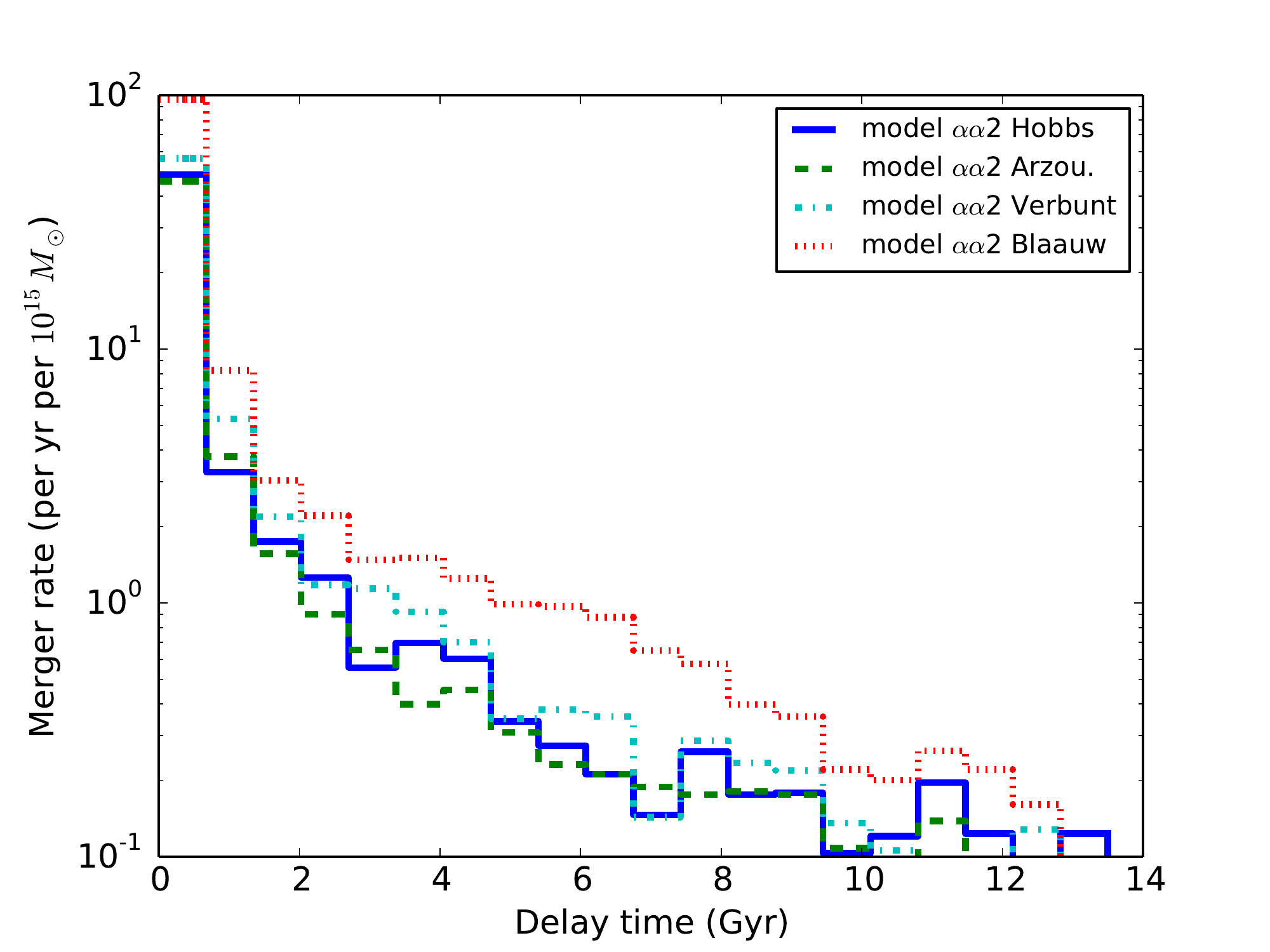}  \\
        \includegraphics[width=\columnwidth]{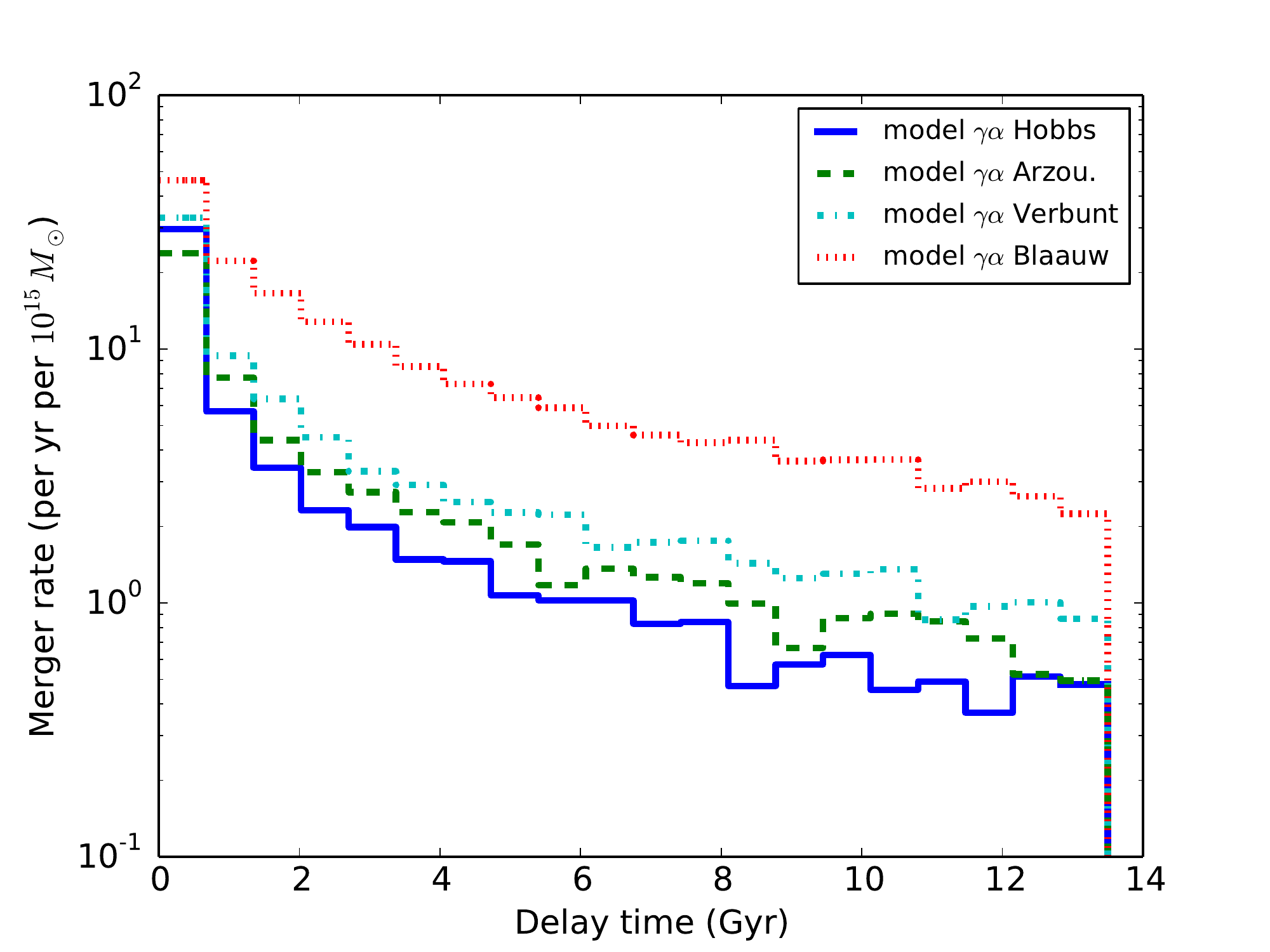}\\
        \end{tabular}
    \caption{Effect of SN-kicks on the delay time distributions of NS-WD mergers for different CE models. The rate is minimized for the large SN kicks of the Hobbs-distribution (blue solid line) and the Arzoumanian-distribution (green dashed line). If the SN kick is only due to the instantaneous mass loss (Blaauw-kick, red dashed line), the rate is maximized. The different panels show the DTDs from different CE models; model \maa\, on top, model \maa2\,in the middle, and model \mga\  on the bottom. See also Fig.\,\ref{fig:DTD_CE} for a direct comparison of the DTDs from the CE models.} 
\label{fig:DTD}
    \end{figure}

The NS-WD merger rate as a function of the delay time (between the merger and formation of the binary with two zero-age MS stars) is shown in Figs.\,\ref{fig:DTD_CE}~and~\ref{fig:DTD}. 
These  DTDs peak at short delay times and then decrease towards longer delay times. The slope of the DTDs differs between different models of the CE phase (Fig.\,\ref{fig:DTD_CE}). The magnitude of the SN kick mostly affects the total number of mergers, and, to a minor degree, the slope of the DTD (Fig.\,\ref{fig:DTD}). Varying the assumptions regarding the stable mass-transfer phase (Sect.\,\ref{sec:mt3}) does not significantly affect the DTD.

The top two panels of Fig.\,\ref{fig:DTD} (see also Fig.\,\ref{fig:DTD_CE}) show very different slopes of the DTD. These represent models \maa\, and \maa2 and indicate that the slope depends strongly on the modelling of the CE phase (line 6 in Fig.\,\ref{fig:wdns_direct}), in particular the efficiency with which orbital energy is used to unbind the envelope. If the CE leads to a more significant shrinkage of the orbit (as in model \maa2), the orbits of the NS-WD binaries tend to be smaller at formation, leading to shorter in-spiral times due to gravitational wave emission. A similar trend is seen in the DTDs of merging massive CO-WDs \citep{Rui09}.

Finally, we turn our attention to those systems with the shortest GW in-spiral times $t_{\rm insp}$. Recently, we showed \citep{Mic16,Mic18} that the in-spiral times of BH-BH, BH-NS, and NS-NS mergers can be as short as years or decades. Therefore, a few $10^{-4}-10^{-1}$ of LIGO gravitational wave sources and short gamma-ray bursts could be preceded by observable SN explosions. The short in-spiral times are achieved in systems that experience a SN kick with the right amplitude and direction, such that the stars are cast into a close and/or highly eccentric orbit. The same mechanism also works for mergers between NS and WDs. With the distributions of non-zero natal kicks, the NS-WD merger rate with in-spiral times less than 100 years is approximately $10^{-7}$\perMsolar, that is, of the order of a few times $0.1\%$ of NS-WD mergers could be preceded by an observable SN explosion years to decades before the merger. For the models with only the mass-loss kick, the in-spiral times of all systems in the current simulations are longer than 100 years. Given our current resolution, this gives an upper limit of approximately $10^{-8}$\perMsolar.

\subsubsection{Characteristics of the white dwarf component}
The composition of the WD component typically consists of carbon and oxygen (Table \,\ref{tbl:rate}). In models \maa~and \mga,\, $\sim$80\% of mergers occur with a carbon-oxygen (CO) WD. In models \maa2, the contribution is reduced to about 60\%. Neutron star mergers with oxygen-neon (ONe) WDs are of secondary importance; contributing about 20\% for models \maa~ and \mga, and 40\% for model \maa2. 

The reason for the increase of ONe WDs around NSs for lower common-envelope efficiencies (such as in model \maa2\, compared to model \maa) is related to the orbital energy available for the progenitors of the WDNS systems. 
In a binary with a high-mass companion compared to a low-mass companion (and other parameters kept constant), there is more orbital energy available. As a result, the orbital shrinkage is weaker, and the CE phase more readily results in a detached binary than in a merger of the two stars. 

Mergers with helium (He) WDs are not frequent in our models. 
As argued in the previous paragraph, NS-WD binaries with less massive companions have less orbital energy and are more likely to merge during the CE phase, which reduces the number of NS-WD with helium WDs. 
The majority of the helium WDs in our simulation form through a different evolutionary channel from those discussed in Sect.\,\ref{sec:ev}. 
For these systems, the initial binary has a large mass ratio and a wide orbit, such that the primary star fills its Roche lobe on the asymptotic giant branch. 
The first phase of mass transfer is unstable in contrast to the other evolutionary channels. Eventually the primary star collapses to a NS. When the secondary star fills its Roche lobe, a last phase of mass transfer commences. Typically this mass-transfer phase is unstable, but if not, a low-mass X-ray binary is formed. After the secondary star has lost its hydrogen envelope, it forms a helium WD. 

Recently, \cite{Zen18} investigated the formation of hybrid WDs, that is, WDs with a carbon-oxygen core and a thick envelope of helium up to $\sim 0.1$\Msolar.  
The existence of a significant mass in helium can catalyse thermonuclear explosions during a merger with another compact object, for example a NS \citep{Met12,Mar16} or a WD \citep[e.g.][]{Pak13}. In the former case, the tidally disrupted WD forms an accretion disk around the NS of mixed He-CO composition and nuclear burning is expected in the disk. In the case of the hybrid WD, however, it proceeds dynamically \citep{Mar16}; further details of such explosions are explored in depth in a forthcoming publication (Zenati et al., in prep.). In our simulations we find that mergers between a NS and a hybrid WD are not common, that is, they represent less than 0.1\% of all mergers. These mergers preferably happen at short delay times ($\lesssim 500 Myr$).

\begin{figure}
    \centering
    \begin{tabular}{c}
        \includegraphics[width=\columnwidth]{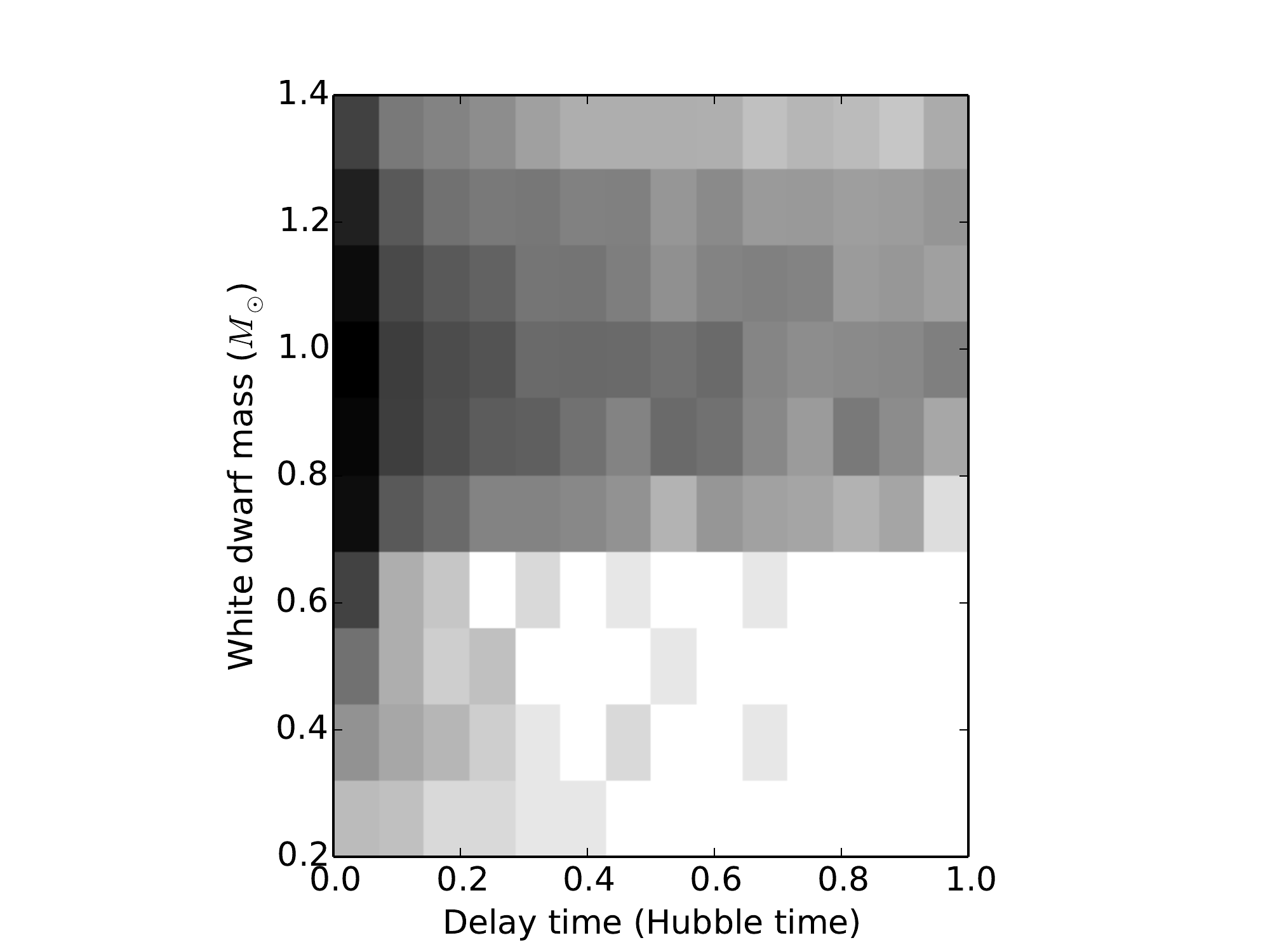}  \\
        \includegraphics[width=\columnwidth]{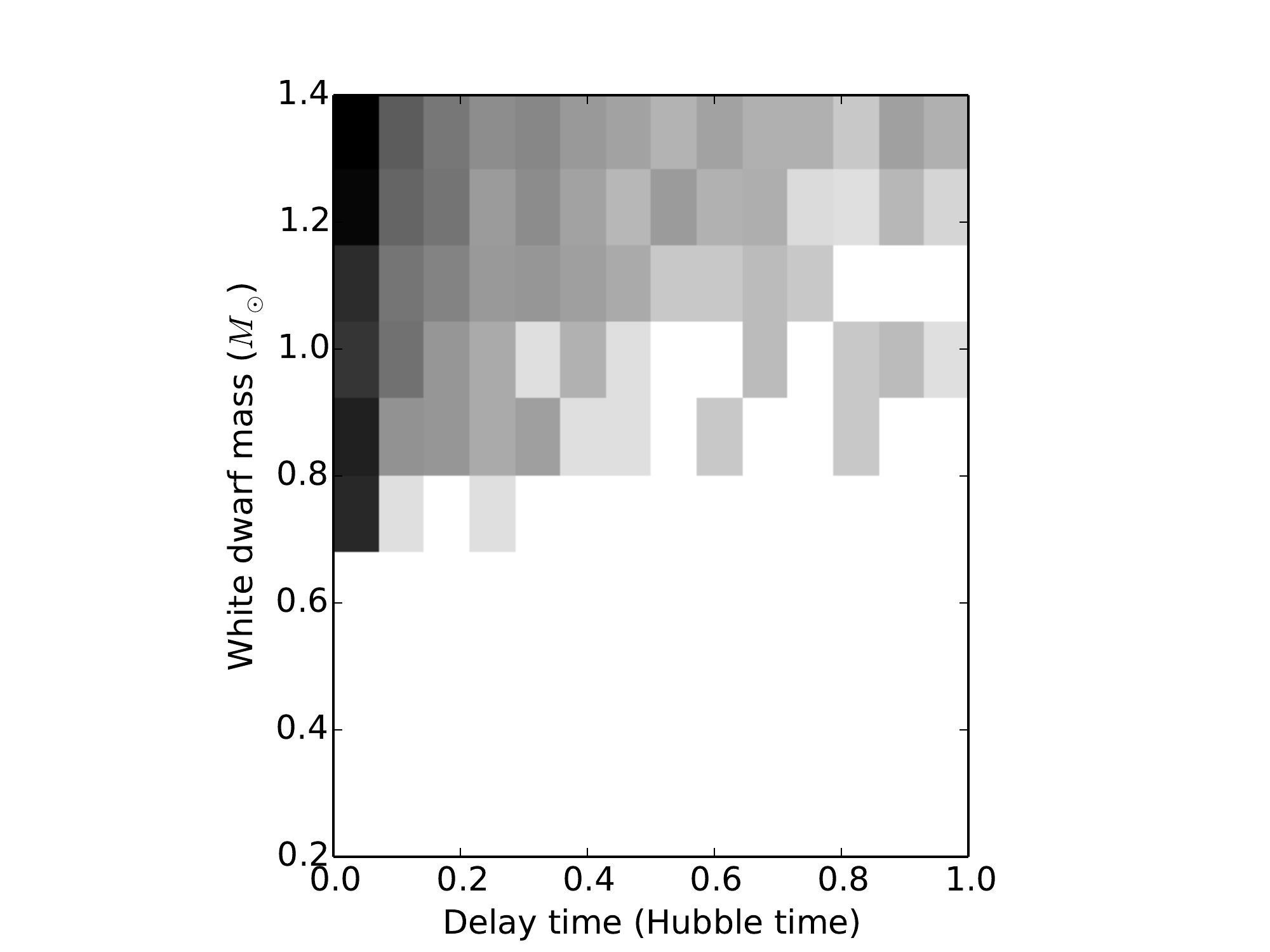}  \\
        \includegraphics[width=\columnwidth]{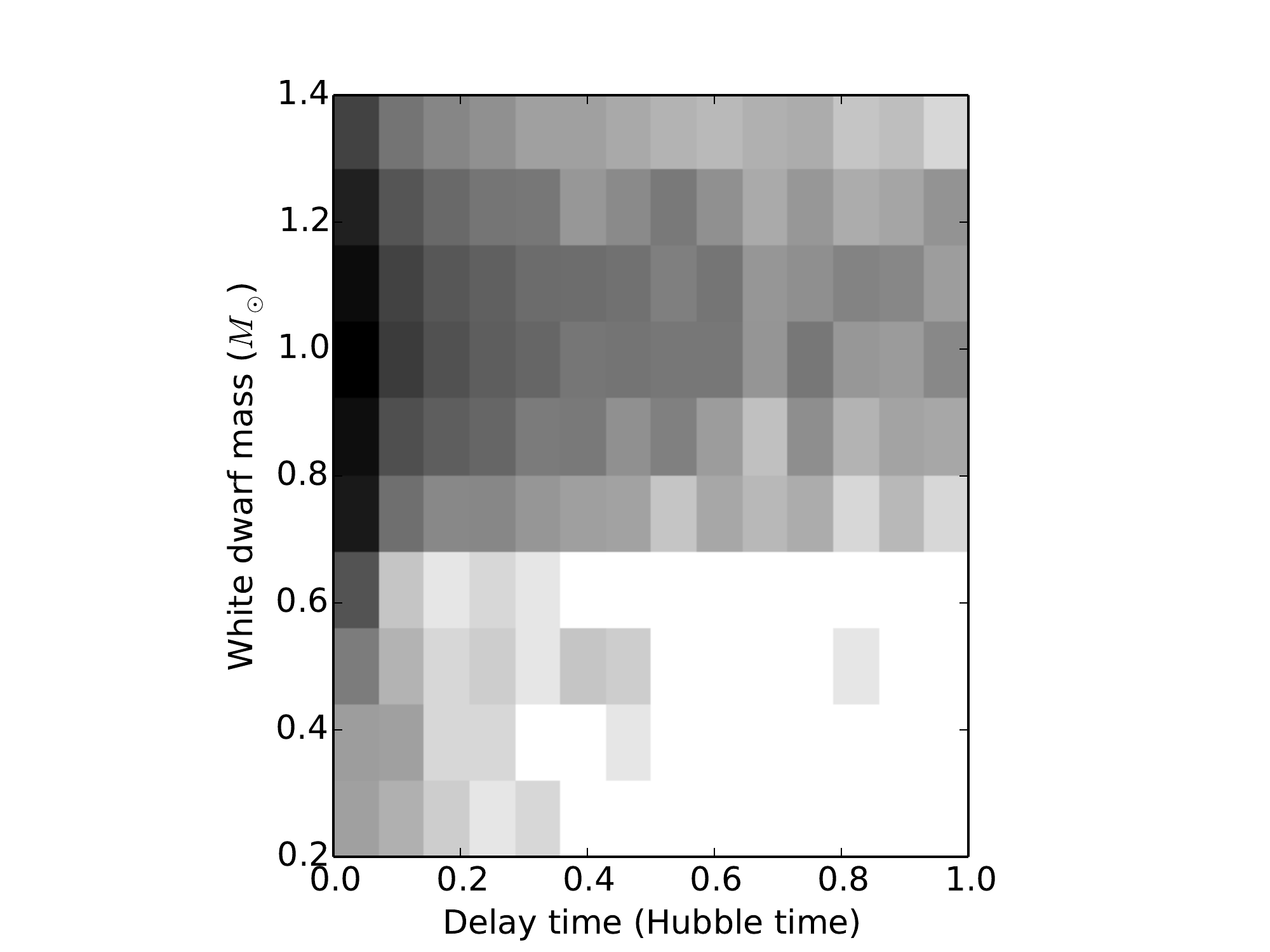}\\
        \end{tabular}
    \caption{White dwarf mass as a function of delay time for three different models of CE evolution. For completeness we show all NS-WD systems that come into contact, not only those that will lead to a merger (Sect.\,\ref{sec:towards_merger}). From top to bottom, the main models \maa, \maa2, and \mga\,are shown, respectively, for the SN kick distribution of \cite{Hob05}. The grey scale corresponds to a density of objects on a linear scale.} 
\label{fig:Mwd}
    \end{figure}

The WD mass as a function of delay time is shown in Fig.\,\ref{fig:Mwd}. In models \maa\, and \mga, we do not find significant evolution of the average WD mass with delay time. For lower CE efficiencies (i.e. model \maa2), the average WD mass can increase somewhat with delay time.

\begin{figure}
    \centering
    \begin{tabular}{c}
        \includegraphics[width=\columnwidth]{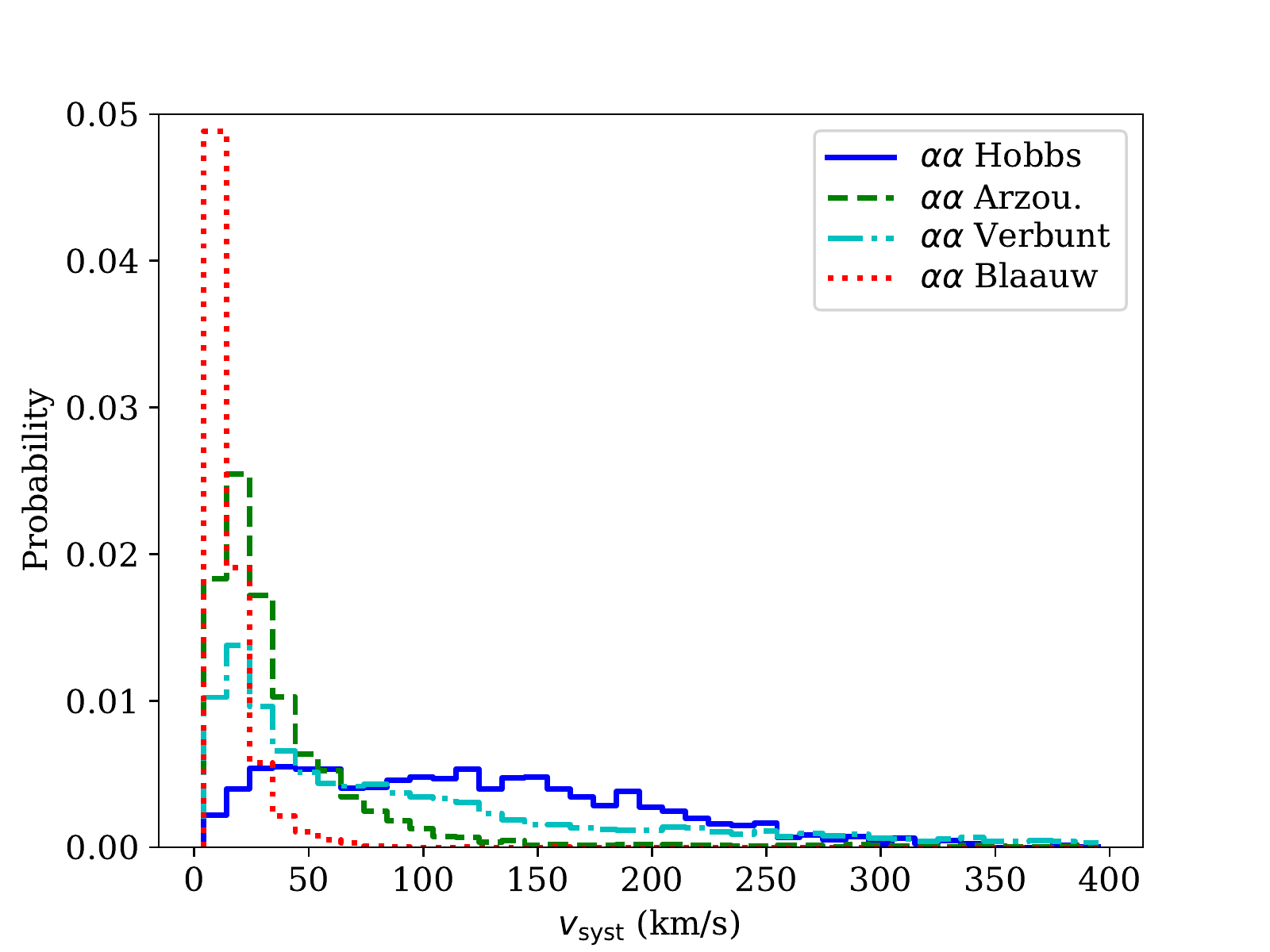}  \\
        \includegraphics[width=\columnwidth]{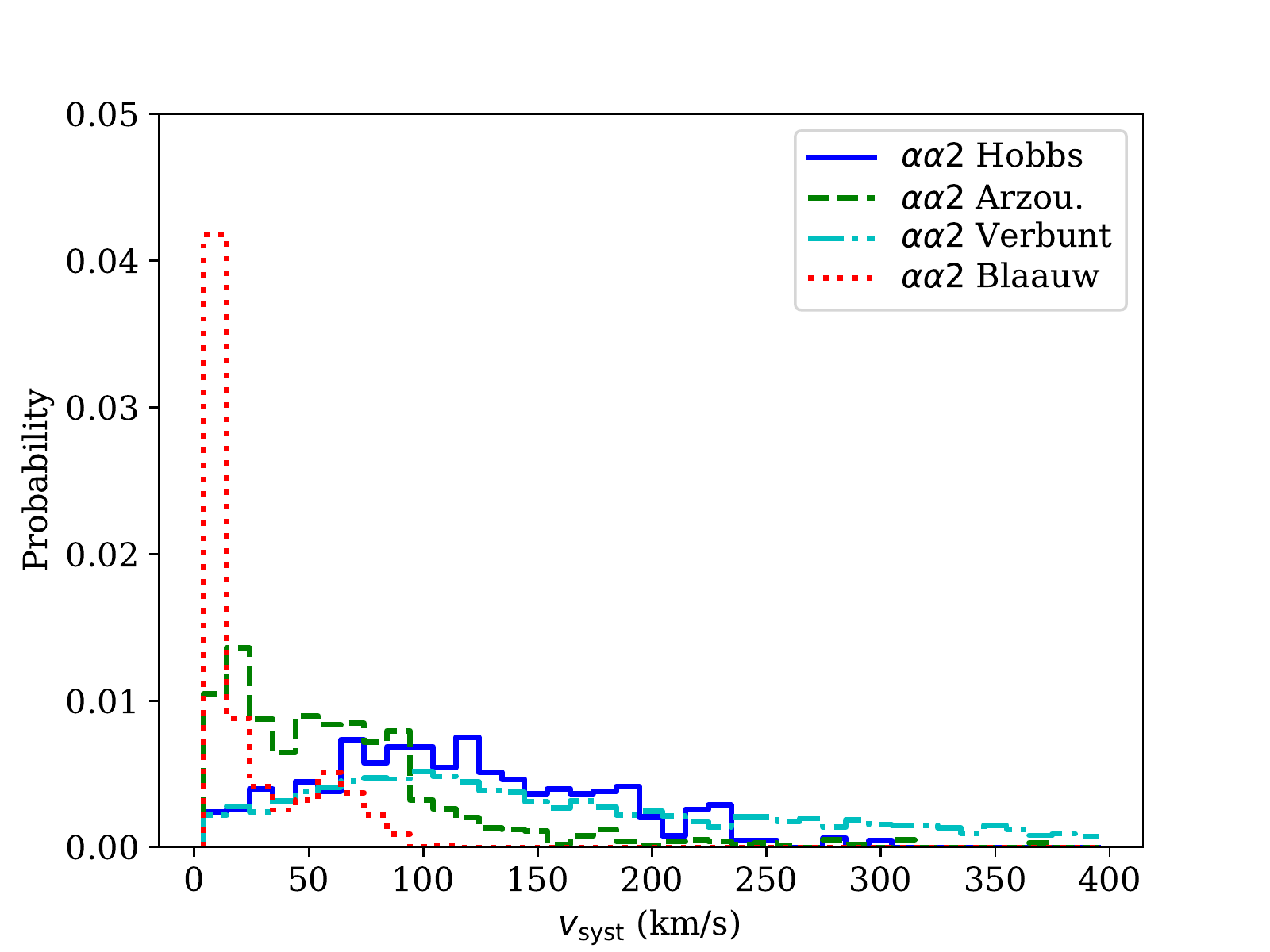}  \\
        \end{tabular}
    \caption{Systemic velocity distribution of NS-WD systems. From top to bottom, the  models \maa and \maa2 are shown.} 
\label{fig:syst_v}
    \end{figure}

 \subsection{Effects of the SN kick}
\label{sec:sn2}

The effect of the SN kick is dependent on the binding energy of the pre-SN orbit and the kinetic energy of the SN kick exerted on the pre-SN orbit.
If the SN kick is large in comparison with the orbital velocity, the system is disrupted.  This happens for example for wide ($a_{\rm SLR} \gtrsim 2\cdot 10^3$\Rsolar) systems in our models with natal kicks. These orbits are sufficiently wide such that the Roche lobe overflow is not expected and the stars live as if they were isolated (not taking into account any effects due to the SN kick). If the mass of the primary stars is in the range of $\sim 8-20$\Msolar, and the secondary mass is in the range of $\sim 1-8$\Msolar, the system could evolve to become a NS-WD binary if the NS kick does not disrupt it. We find that the formation rate of disrupted binaries with one NS component and one WD component after a Hubble time is about $10^{-3}$\perMsolar\, if natal kicks are taken into account. This is about two orders of magnitude above the time-integrated NSWD-merger rate in the same models. For lower SN kicks (i.e. our models without a natal kick, and only a mass-loss kick), the disruption rate is $5\cdot 10^{-4}$\perMsolar. 

If the magnitude of the SN kick is of the same order as the average orbital velocity, the post-SN stellar orbits are dependent on the direction of the SN kick and the orbital phase for eccentric orbits. If the stars are at apocenter and the SN kick is aligned with the orbital velocity, the binary disrupts readily. On the other hand if the SN kick is anti-aligned, the orbit shrinks. This can give rise to NSWD binaries that merge on short timescales after the SN explosion (Sect.\,\ref{sec:dtd}), and to systems in which only the secondary star initiates a phase of mass transfer. The latter systems are  initially sufficiently wide, such that the primary star does not fill its Roche lobe, and the stars evolve as if they were single isolated stars until the primary collapses in a SN explosion, in which the orbit shrinks. From this point on, the system evolves in a similar fashion as in the  `NS-WD direct' channel (see Fig.\,\ref{fig:nswd_direct} from line 5 on wards); the secondary initiates one or two mass-transfer phases, and after the formation of the NSWD, the system merges due to the emission of gravitational waves. The time-integrated rate of this channel is a few $10^{-8}$\perMsolar.

\subsection{The offset distribution}
\label{sec:offset}
The SN kick imparts some velocity to the binary even if the NS receives no natal kick; see Figure~\ref{fig:syst_v} for the Blaauw kick. Therefore, the spatial distribution  of mergers can differ from the initial galaxy density distribution, that is, the merger takes place at an offset. This effect can be used to identify WD-NS mergers among the bulk of transient extragalactic events, and eventually even to  distinguish between different natal kick distributions for NSs. 

In order to estimate the expected offsets for the NS-WD mergers of our BPS models, we integrate the motion of each binary in various galaxy potentials. The binary motion is integrated starting from the formation of the NS until the final merger (using the python package \texttt{galpy}\footnote{http://github.com/jobovy/galpy} \citep{Bov2015} which makes use of fourth-order symplectic integrator). Below we briefly describe the setup of these simulations and the resulting distributions of offsets.

Due to the wide delay time distribution of NS-WD mergers, these events can take place in both old (elliptical) and young (disk) host galaxies. The gravitational field of the host and its symmetry play an important role in establishing the offset distribution. 
We consider four types of galaxies
\cite[based on][]{Bel2002,Bel2006}: dwarf and giant ellipticals, as well as typical and bulgeless disk galaxies. The giant elliptical is constructed out of two potentials, one for the stellar and one for the dark matter component. For the former we adopt the Hernquist potential \citep{Her90} with a scale length of $a=5$~kpc and mass $\mathcal{M}=5\times 10^{11}$~M$_\odot$, and a Navarro–Frenk–White (NFW) potential \citep{Nfw96} with the same parameters. The model for the dwarf elliptical galaxy contains the same two potentials but with $a=0.5$~kpc and $\mathcal{M}= 5\times 10^8$~M$_\odot$. 
The disk galaxy is constructed out of three potentials; 1) a Miyamoto-Nagai potential \citep{MNP1975} for the disk taking a radial scale length of $a_d = 4.2$~kpc, a vertical scale length of $b_d = 0.198$~kpc, and mass $\mathcal{M} = 8.78\times 10^{10}$~M$_\odot$,
2) a Hernquist potential for the bulge with $a=0.79$~kpc and $\mathcal{M} = 1.12\times 10^{10}$~M$_\odot$, and 
3) a NFW potential for the dark matter halo with $a = 6$~kpc, $\mathcal{M} = 5\times 10^{10}$~M$_\odot$ and a cutoff at 100kpc. 

The initial spatial distribution of the binaries follows the Hernquist density profile in the case of the elliptical galaxy. For the disk galaxy, the initial positions of the stars are drawn from the Miyamoto-Nagai density profile (disk origin) or the Hernquist density profile (bulge origin). 
The initial speed (before the kick is imparted due to SN explosion) of the disk-born binary is chosen to be  the circular speed. 
For the initial speeds in the bulges, we follow the prescription from \cite{Bel2002,Bel2006}, that is, we use the circular speed and randomise the orientation of the angular momentum. We have tested an alternative approach, where we draw the initial speed  using the stellar distribution function obtained under the assumption that the stellar system is ergodic and has reached a dynamical equilibrium. Our simulations show that these two approaches lead to very similar offset distributions.  

We take the star formation history of the different types of galaxies into account. 
We assume a constant star formation rate (SFR) for 10 Gyr for the disk galaxy without a bulge. In addition, for the case of the disk galaxy with a bulge, we assume star formation is enhanced in the bulge by a factor 10 during the first gigayear. For elliptical galaxies, we assume a constant SFR for the first 5 Gyr. 

We assume that the orientation of the recoil velocity has a uniform distribution on a sphere. The recoil velocity (see Figure~\ref{fig:syst_v} for the distribution) is added to the binary velocity in the galaxy as a vector. 
The energy conservation level of the integrations is checked for each individual orbit to remain below $|(E(t_\mathrm{merger}) - E_0)/E_0| < 10^{-6}$. If the energy is not well conserved, the orbit is integrated once again with  time steps that are 
80 times shorter and checked for energy conservation on level $|(E(t_\mathrm{merger}) - E_0)/E_0| < 10^{-4}$. If the integrator still cannot guarantee the energy conservation, the orbit is removed from consideration. Usually,  3-5 of the $\sim 10000$ orbits are removed, which should not significantly affect our results.

A comparison of the recoil speeds plotted in Figure~\ref{fig:syst_v} already demonstrates a few interesting features: in the case without a natal kick, the binary can reach a speed up to 70~km/s and most of the binaries move with a speed of $\approx 30$~km/s. For all natal kick distributions we see slightly higher systemic speeds in the $\alpha \alpha 2$ model of common-envelope evolution compared to the $\alpha\alpha$ model, as the pre-SN orbits tend to be smaller in the prior model.
The natal kick distribution suggested by \cite{Hob05} gives rise to a broad velocity distribution which peaks at $\approx 100$~km/s. The \cite{Arz02} kick transforms into a narrow distribution of recoil speeds with a peak at $\approx 40$~km/s and nearly a zero fraction at speeds $v>200$~km/s. 
The systemic velocity distribution derived for the natal kicks as derived by \cite{Ver2017} peaks at $\approx 40$~km/s and decays slowly till high velocities, disappearing completely by $\approx 300$~km/s ($\alpha \alpha$ model). In the case of the $\alpha \alpha2$ model, the peak is at $\approx 100$~km/s and the tail of speed distribution continues until $\approx 400$~km/s.

We find that binaries receiving recoil speeds of $\sim 200$~km/s escape from dwarf elliptical galaxies, whereas they remain bound in the case of giant elliptical hosts. The offset from the centre of a giant elliptical galaxy is negligible, so the spatial distribution of mergers follows the same stellar density of the underlying giant elliptical host galaxies.

For a disk galaxy the stellar density decays faster in the vertical direction, so a binary with a recoil speed of $\sim 100$ km/s in this direction can be seen at a noticeable offset (1-10 kpc) when viewed edge-on, but not in the case of face-on orientation. Our simulations show that the offset distribution is very similar for the case of the disk galaxy with and without a bulge.

\begin{figure*}
    \centering
    \begin{tabular}{cc}
        
\includegraphics[width=\columnwidth]{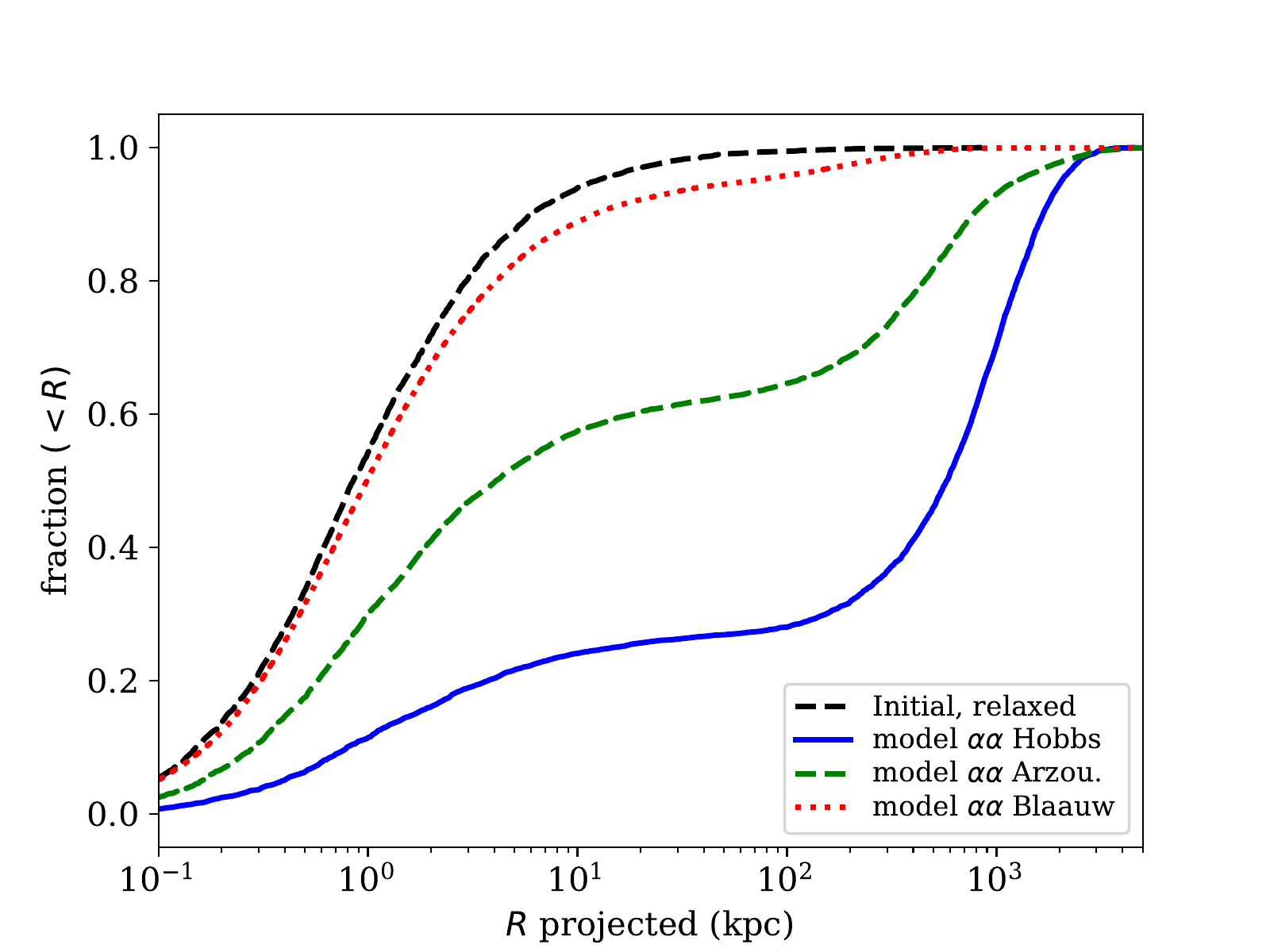} & 
\includegraphics[width=\columnwidth]{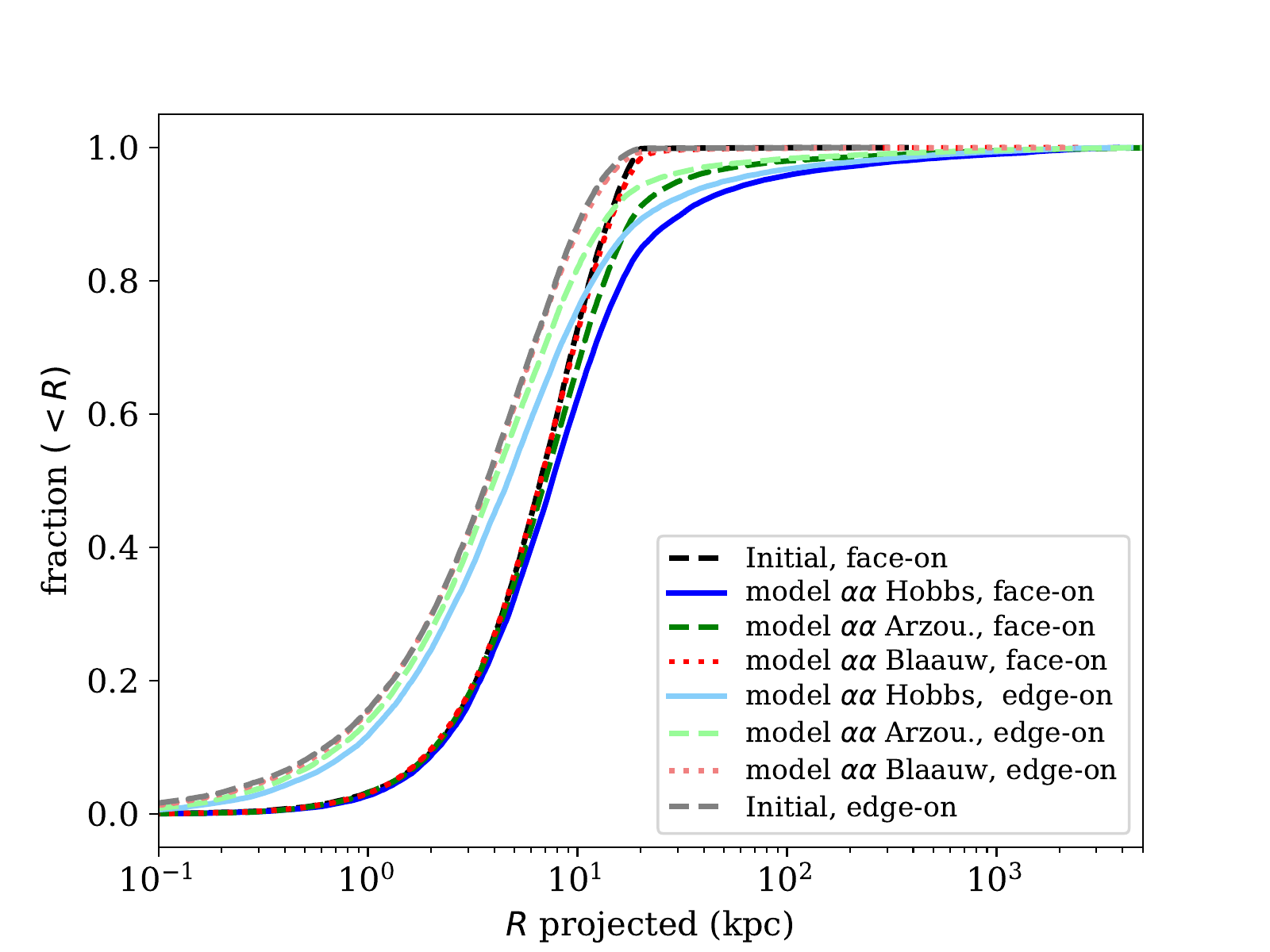} \\
\includegraphics[width=\columnwidth]{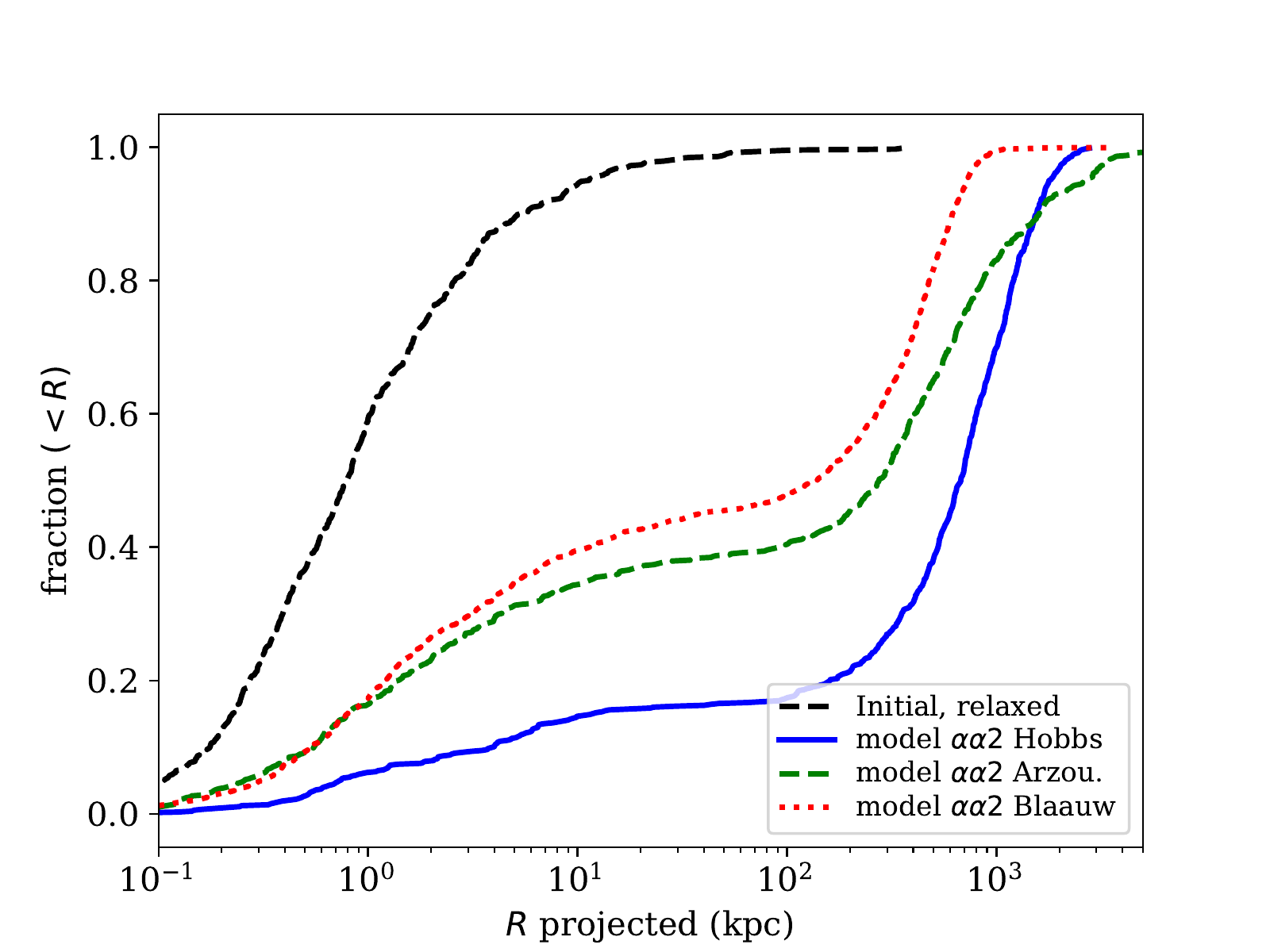} &
\includegraphics[width=\columnwidth]{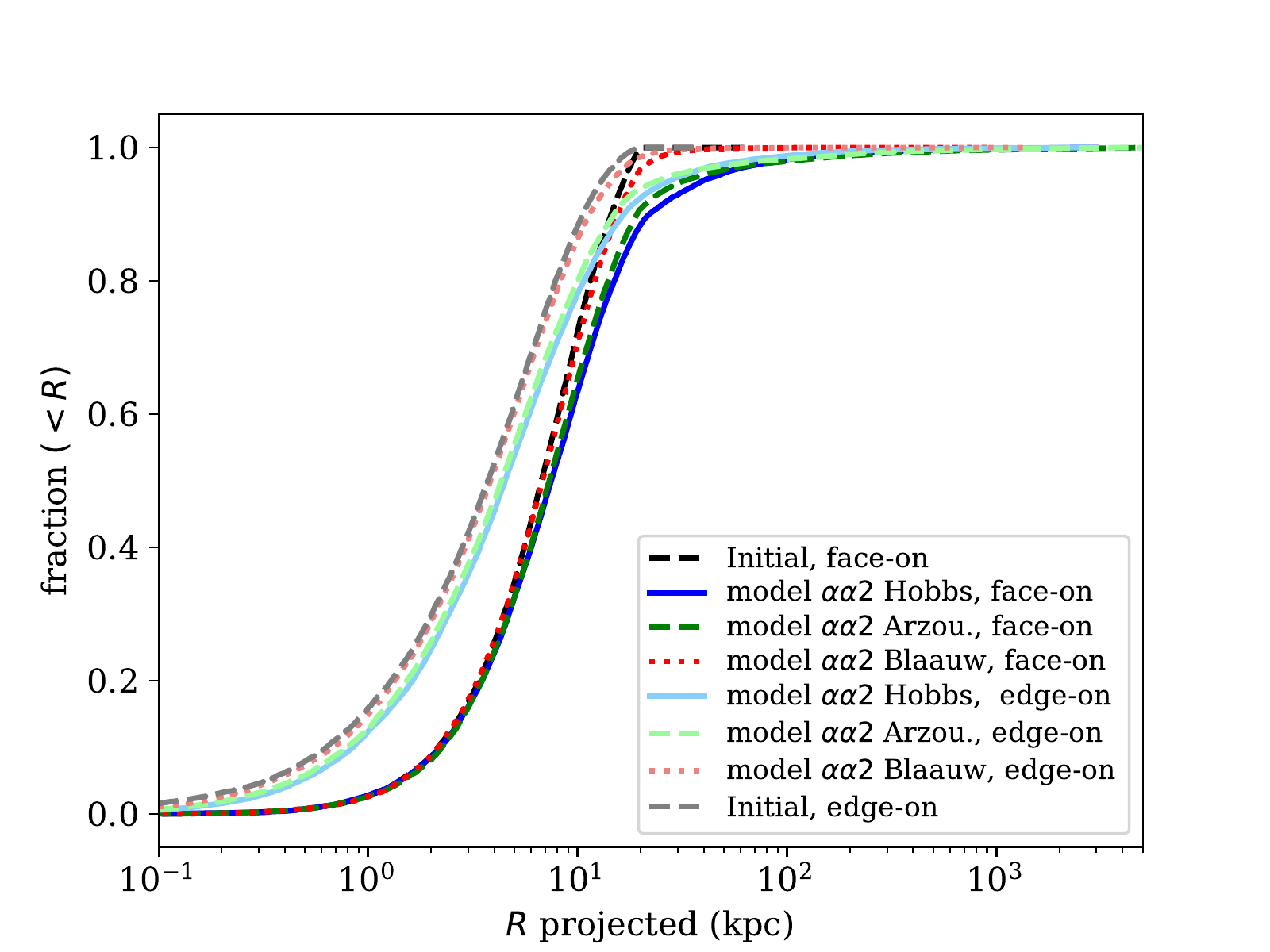} \\
        
        \end{tabular}
    \caption{The cumulative distribution of offsets from the centre of the dwarf elliptical (left column) and the disk galaxy with bulge (right column). The upper row shows results for the CE model $\alpha\alpha$ and the lower row for the CE model $\alpha\alpha2$ . The offset distribution using the kicks of \citet{Ver2017} is very similar to that using the kick distribution of \citet{Arz02}. } 
\label{fig:offsets}
    \end{figure*}

In Figure~\ref{fig:offsets}, we compare the initial and final (at the time of the merger) offset distributions for the dwarf elliptical and the disk galaxy models. 
The general trend is that while larger kicks reduce the number of mergers, they give rise to larger offsets from the host galaxy. For the disk galaxy, the mean offset (face-on orientation) is $\sim$20-40 kpc for the strong natal-kicks of the Hobbs distribution, $\sim$20-30 kpc for the Arzoumanian and Verbunt distributions, and $\sim$8 kpc without natal kicks. For the elliptical galaxy, the mean offsets are $\sim$700-800 kpc, $\sim$250-600 kpc,  $\sim$250-500kpc, and $\sim$20-250 kpc, for the Hobbs-, Arzoumanian-, Verbunt-, and Blaauw distributions, respectively. The offsets are most pronounced in the dwarf galaxy; over 40\% of systems are found at offsets around 0.1-1 Mpc for the Arzoumanian distribution, and this increases to 80\% for the Hobbs distribution.

\section{Discussion and summary}
\label{sec:dis}
In this work we have made the first systematic study of the demographics of NS-WD mergers, their properties, rates, delay-time distributions and observable aspects. We have considered a variety of initial conditions and studied the sensitivity of our results to the various uncertainties in the physical processes involved in the stellar evolution of NS-WD progenitors. In particular we have explored the dependence on NS natal kicks, mass transfer in binaries, and common-envelope evolution.   

{\bf Evolutionary channels:} We find that in the main channel for NS-WD mergers, the white dwarf forms before the NS.  The primary transfers mass to the secondary as it evolves off the main sequence, increasing the mass of the  secondary. The latter then becomes more massive than the original primary and ends its life as a neutron star after the primary has already become a WD. 

The overall results depend on the various assumptions and initial conditions. However, the different models we consider provide qualitatively similar predictions, and quantitatively differ by at most a factor of two to three in the overall expected time-integrated rates. Tables \ref{tbl:rate}-\ref{tbl:rate_obs} summarise the main differences.   

{\bf Rates, delay-time distributions, and host galaxies:} 
The time-integrated rates of NS-WD mergers depend on the specific assumptions and initial conditions considered, but are generally in the range of $(3-7)\cdot 10^{-5}$ \perMsolar. If reduced kicks are common among the progenitors of NS-WD, their merger rate increases moderately; under the extreme assumption of a minimal SN kick (i.e. no natal kick) for all systems, the upper limit to the merger rate is only slightly larger at $\lesssim 20\cdot 10^{-5}$ \perMsolar.

For the Milky Way, assuming a constant SFR of 3\Msolar/yr, we generally find a merger rate of $1-2\cdot 10^{-4}$ per year (up to $8\cdot 10^{-6} -5\cdot 10^{-4}$yr$^{-1}$ for the full range of models). This is in good agreement with the rate of $1.4\cdot 10^{-4}$yr$^{-1}$ based on other binary population synthesis calculations \citep{Nel01}, and $2.6\cdot 10^{-4}$yr$^{-1}$ based on binary radio pulsars \citep{Bob17}, but slightly larger than the rate of $(1-10)\cdot 10^{-6}$yr$^{-1}$ and $\sim 4.1\cdot 10^{-6}$yr$^{-1}$ based on statistics of observed low-mass X-ray binaries \citep{Coo04} and pulsar binaries with WD companions \citep{Kim04}, respectively. It is unclear where the difference in the observational estimates comes from; larger samples and a better understanding of observational selection effects are necessary to assess if the theoretical estimates are at odds with nature.

The observed time-integrated rate of supernova type Ia in field galaxies is about $10^{-3}$\perMsolar \citep[e.g.][]{Mao14, Mao17}.  
The predicted rate of NS-WD mergers is about $2-6\%$ (up to $14\%$) of the observed type Ia SN rate, and about $1-6\%$ of the predicted rate of WD-WD mergers 
(see Tables \ref{tbl:rate}-\ref{tbl:rate_obs}).

The delay time distribution peaks at early times ($<1-2$ Gyrs), but the tail distribution extends up to a Hubble time. In a small fraction (a few times $0.1\%$) of the cases, the SN producing the NS precedes the NS-WD merger by less than 100 yrs, and is therefore potentially observable. The delay time distribution peaking at early times suggest that NS-WD mergers are most likely to be found in late-type, disk, and star-forming galaxies with only small fractions expected to be found in early type elliptical/S0 galaxies.     

{\bf Composition:} In most models the majority of NS-WD mergers involve CO-WDs with smaller fractions of ONe, and very small fractions of He WDs ($0.3-1.4\%$; and a negligible fraction of hybrid He-CO WDs).

{\bf Offsets:} The offsets of the expected location of NS-WD mergers generally follow the stellar density of their host galaxies. The only exception is the case of dwarf galaxies whose escape velocity is small. In models which include NS natal kicks, the amplitude of the kicks could be sufficiently large as to eject the NS-WD binaries from the host galaxy, leading to very large offsets  of up to a few hundred kiloparsecs.

{\bf Possible observational candidates:} Given the predicted properties, one may consider the observational manifestation of NS-WD mergers. 
\citet{Met12} suggested that NS-WD merges could be related to the class of faint type Ib Ca-rich SNe \citep{Per10}. However, our results indicate that the DTD of NS-WD mergers is inconsistent with the observed distribution of Ca-rich SNe, which typically explode in old environments \citep{Per10,Kas12,Lym13,Per14}. Given the WD compositions (lacking in helium WDs), and the small amounts of intermediate and iron elements produced in such explosions (Zenati et al., in prep.), these transients are likely most similar, spectrally, to type Ic SNe. This, again, is at odds with the observed helium and the significant amounts of intermediate elements in the spectra of Ca-rich SNe \citep{Per10}. The overall time-integrated rates we find could be consistent with the estimated rates of Ca-rich SNe \citep{Per10}, but more recent estimates based on the PTF survey data suggest that the rates of Ca-rich SNe could be significantly higher ($33-94\%$ of the Ia SNe rate \citep{Fro18};  $3-7$ times the highest rate estimated in any of our models, Table \ref{tbl:rate_obs}). 

Another possibility is that NS-WD mergers manifest observationally as rapidly evolving SNe, a class, or several classes, of rapidly declining and energetically weak likely type-Ic SNe \citep{Che88,Poz10,Kas10,Per11,Dro13, Dro14} mostly exploding in late-type galaxies. The estimated rates and DTD of such SNe could be consistent with our demographic results of NS-WD mergers, but more detailed studies of the expected light curves and spectra of such explosions are required for direct comparison. 

\begin{table*}[h]
\caption{
Comparison of our synthetic merger rates with observed transient rates, (see also Table\,\ref{tbl:rate}). The synthetic merger rates comprise those of NS-WD systems, WD-WD systems, and a subset of the latter in which only mergers between carbon-oxygen WDs with a combined mass above the Chandrasekhar mass are taken into account (i.e. the classical double degenerate channel for SNIa, \citealt{Web84,Ibe84}). For NS-WD mergers a range of rates is shown based on different models for the SN kick.
}
\begin{tabular}{|l|ccc|}
\hline
 & \multicolumn{3}{c|}{Synthetic time-integrated merger rate (\perMsolar)}\\
\hline
BPS Model &   NS-WD  & WD-WD & Super-Ch WD-WD \\
\hline
\hline
\maa & $(4.7-18) \cdot 10^{-5}$ &   $3.2\cdot 10^{-3}$  & $5.5\cdot 10^{-4}$  \\
\maa2 & $(3.8-8.1) \cdot 10^{-5}$&  $5.9\cdot 10^{-5}$  & $2.1\cdot 10^{-5}$  \\
\mga &  $(3.7-12) \cdot 10^{-5}$ &  $3.2\cdot 10^{-3}$  &  $4.2\cdot 10^{-4}$ \\
\hline
\hline
 & \multicolumn{3}{c|}{Observed time-integrated rate ($$\perMsolar)}\\
\hline
Supernova Type Ia$^2$  & \multicolumn{3}{l|}{$(1.3\pm0.1)\cdot 10^{-3},\, (1.6\pm0.3)\cdot 10^{-3} $} \\
Calcium-rich transients$^1$ & \multicolumn{3}{l|}{33-94\% of the SNIa rate}\\
\hline
\end{tabular}
\label{tbl:rate_obs}
\begin{flushleft}
\tablefoot{ $^1$ \citet{Mao17};$^2$ \citet{Fro18};}
\end{flushleft}
\end{table*}

\begin{acknowledgements}
ST thanks Ruben Boots and Manos Zapartas for the software to make the Roche lobe diagrams.
ST acknowledges support from the Netherlands Research Council NWO (grant VENI [nr. 639.041.645]). 
HBP \& AI acknowledges support from the Israel science foundation I-CORE program 1829/12.
\end{acknowledgements}

\bibliographystyle{aa}
\bibliography{bibtex_silvia_toonen.bib}

\end{document}